\documentclass[useAMS,usenatbib,onecolumn]{mn2e}
\usepackage{amsmath}
\usepackage{graphicx}
\usepackage{amssymb}
\usepackage{sidecap}
\usepackage{color}

\def\dalemb#1#2{{\vbox{\hrule height.#2pt
        \hbox{\vrule width.#2pt height#1pt \kern#1pt \vrule width.#2pt}
        \hrule height.#2pt}}}

\def\ba{\begin{eqnarray}}
\def\ea{\end{eqnarray}}
\def\be{\begin{equation}}
\def\ee{\end{equation}}

\def\gtorder{\mathrel{\raise.3ex\hbox{$>$}\mkern-14mu
             \lower0.6ex\hbox{$\sim$}}}
\def\ltorder{\mathrel{\raise.3ex\hbox{$<$}\mkern-14mu
             \lower0.6ex\hbox{$\sim$}}}

\title%
[Filling in CMB map missing data using constrained Gaussian realizations]%
{Filling in CMB map missing data using constrained Gaussian realizations}

\author[Bucher \& Louis]{
Martin Bucher 
${}^{1}$\thanks{E-mail: bucher@apc.univ-paris7.fr} and
Thibaut Louis
${}^{1,2,3}$\thanks{E-mail: thibaut.louis@ens-lyon.fr}\\
${}^{1}$%
Laboratoire Astroparticule \& Cosmologie, 
Universit\'e Paris Diderot 7, 
10 rue Alice Domon et L\'eonie Duquet,
75013 Paris, France\\
${}^{2}$Laboratoire de Physique,
ENS de Lyon,
46, all\'ee d'Italie
F69007 Lyon, France \\
${}^{3}$%
Oxford Astrophysics, Department of Physics, Denys Wilkinson Building, Keble Road, Oxford OX1 3RH, UK 
}

\begin{document}

\date{1 September 2011; revised 11 September 2011}


\label{firstpage}

\maketitle

\begin{abstract}
For analyzing maps of the cosmic microwave background sky, it is
necessary to mask out the region around the Galactic equator where the
parasitic foreground emission is strongest as well as the brightest
compact sources. Since many of the analyses of the data, particularly
those searching for non-Gaussianity of a primordial origin, are
most straightforwardly carried out on full-sky maps, it is of great interest to
develop efficient algorithms for filling in the missing information
in a plausible way. In this paper we explore algorithms for 
filling in based on constrained Gaussian realizations. Although
carrying out such realizations is in principle straightforward, for
finely pixelized maps as will be required for the Planck analysis
a direct brute force method is not numerically tractable. Here
we present some concrete solutions to this problem, both on a 
spatially flat sky with periodic boundary conditions
and on the pixelized sphere. One 
approach is to solve the linear system with an appropriately preconditioned
conjugate gradient method. While this approach was successfully implemented
on a rectangular domain with periodic boundary conditions and worked
even for very wide masked regions, we found that the method 
failed on the pixelized sphere for reasons that we explain here.
We present an approach that works for full-sky pixelized maps on
the sphere involving a kernel-based multi-resolution Laplace
solver followed by a series of conjugate gradient corrections
near the boundary of the mask. 
\end{abstract}

\section{Introduction}

The analysis of the anisotropies both in temperature and polarization of the primeval cosmic
microwave background has contributed substantially to our present understanding of the Universe
on the largest scales and at the earliest times and continues to be an area of intense activity.
According to the simplest inflationary models, the primordial cosmological perturbations should
be very nearly Gaussian. Consequently much of the focus has been on exploiting the power spectrum,
 which for an isotropic Gaussian underlying statistical process would
encapsulate all the exploitable information. Initial studies of CMB maps indeed show that any
non-Gaussianity that may be present in the primordial signal must be very small. Nevertheless tests
of non-Gaussianity provide a powerful test of the current cosmological paradigm and of possible new 
physics, and inflationary models do exist predicting levels of non-Gaussianity that will be measurable
with the Planck space mission and other new data sets.

One of the obstacles to characterizing the primordial CMB anisotropies is the emission from Galactic
and extragalactic foregrounds. Indeed how best to separate and remove these foregrounds is an area
of intense activity. However near the Galactic plane and in the immediate vicinity of strong point
sources, the best strategy is to mask these regions, giving sky maps with a cut around the Galactic
equator of varying width and numerous holes. Theoretical models are best analyzed in terms of 
full-sky maps, so given this situation, there are two alternatives: (1) one can adapt estimators to
cut sky maps, although doing so often involves linear algebra problems that are ill-conditioned and
numerically extremely demanding, (2) one can
fill-in the maps with various algorithms to obtain a plausible full-sky map
and then analyze these using full-sky algorithms. The philosophy behind approach (2) is somewhat suspect 
and studies, for example by Monte Carlo, of the effect or possible artifacts of the filling in algorithm 
are required. Nevertheless, if the underlying statistical process is not far from Gaussian and the fraction 
of the sky masked out is small, one would expect the error to be minimal and characterizable. 

Several algorithms have been proposed to fill-in maps. (This process is sometimes called `in-painting.')
\cite{Abrial,AbrialB,perotto} for example have proposed using techniques based on sparsity that have proved remarkably
successful for restoring images from photographs that have been blurred and have a large number of missing pixels. This
technique has the advantage that it does not require any assumptions regarding the underlying power spectrum.
However one drawback of this approach is that it is difficult to know whether these techniques might 
introduce parasite non-Gaussianity not already present in the original maps. In this paper we instead consider
filling in by means of constrained Gaussian realizations. Under this approach it is assumed that the power
spectrum can be characterized sufficiently well based on the data from the unmasked pixels. Then a random map is generated
taking the observed pixels as given. Let us denote a sky map by the vector in block form 
${\bf z}=({\bf x}, {\bf y})$ where ${\bf x}$ corresponds to those fixed which have been masked, or not observed, 
and ${\bf y}$ corresponds to those pixels that have been observed, or not masked. We assume that
the joint distribution is given by a Gaussian distribution 
$P({\bf z})=(2\pi )^{-n_{pix}/2}{\rm det}^{-1/2}[{\bf C}] \exp \left[-\frac{1}{2} {\bf z}^T {\bf C}^{-1} {\bf z} \right] $
and wish to fill-in based on the conditional probability distribution 
$P({\bf x}\vert {\bf y}).$ Formally the problem may be solved almost trivially using 
elementary matrix algebra and the solution reduces to the form
$P({\bf x}\vert {\bf y})=(2\pi )^{-n_{masked}/2}\det ^{-1/2}[{\bf C}_{constr}]
\exp \left[ -(1/2)
({\bf x}-{\bf x}_{ML})^T {\bf C}_{const}^{-1} ({\bf x}-{\bf x}_{ML})
\right] $ 
where ${\bf C}_{const}={\bf C}_{\bf xx}^{-1}$ and 
${\bf x}_{ML}= (({{\bf C}^{-1}})_{\bf xx})^{-1}({\bf C}^{-1})_{\bf xy}{\bf y}.$
The maximum likelihood map in a cut region is essentially a Wiener filter as discussed
in the context of CMB maps by \cite{bunn,bunnc,bunnb}. (For a more recent discussion
of some related issues see \cite{feeney}.)
If the number of pixels in a map were modest, a direct solution would be trivial. 
However for CMB maps of the resolution needed for Planck, one has typically up to 
$5\times 10^7$ pixels (corresponding to a Healpix map with $n_{side}=2048),$ so
operations of order $n_{pix}^2$ are to be avoided if at all possible, and operations
of order $n_{pix}^3$ such as would be required to solve a linear system directly or to invert
a matrix are ruled out. While ${\bf C}^{-1}$ may take a particularly simple form over the
full sky, for example when both the underlying cosmological model and the instrument noise
are isotropic, the presence of the projection operators onto the masked and unmasked regions
breaks this isotropy, introducing nonzero off-diagonal elements almost everywhere.
Moreover because of the nearly scale invariant nature of the underlying spectrum, the matrices
are poorly conditioned.

There is some overlap between this work and studies of
Gibbs sampling for CMB power spectrum estimation
as discussed for example in \cite{wandelt_one,ericksen,ericksen2}.
Here one of the steps in the
Gibbs sampling loop is to generate a map for the underlying
CMB anisotopropies $s$ according the conditional distribution
$P(s\vert P, m)$ where $P$ is the assumed power spectrum
and $m$ is the maximum-likelihood map (resulting as
an output from a map maker program) whose deviations from $s$
are described by the inverse noise matrix $N^{-1}.$ Under this
approach foregrounds may either by incorporated into $N^{-1}$
or $(N^{-1})_{aa}$ (where $a$ is the pixel index)
may be set to zero, which has the effect
of masking out those pixels. Another related filling is approach
called `harmonic inpainting' \cite{inoue}. This last approach
is somewhat ad hoc but has been shown to yield reasonable results.

Some issues associated with constrained Gaussian realizations have been
discussed in \cite{Cardoso}. We are aware of a number of private codes
for carrying out constrained Gaussian realizations on the sphere, in
many cases only with coarser pixelizations. However at present no systematic    
discussion of this problem seems to exist in the
literature. This paper endeavors to fill this gap. We found the
related papers of \cite {Hoffman91} and \cite{Zaroub94} useful.

The organization of the paper is as follows. In Section 2 we explore a solution based on the 
conjugate gradient method. The problem as formally presented above can be divided into
two operations: (1) given a ${\bf y}$ calculating ${\bf x}_{ML},$
and (2) producing a random vector $\delta x$ whose covariance matrix is ${{\bf C}^{-1}}_{\bf xx}.$
Operation (2) is not necessarily trivial when the dimension is large and 
${{\bf C}^{-1}}_{\bf xx}$ has a lot of diagonal elements. We show how using a trick due to 
\cite{Hoffman91},
instead of carrying out (1)+(2), the procedure can be simplified by carrying out 
a random realization over the whole sky, followed by (1) acting on ${\bf y}$ plus another component. 
It is shown how (1) can be pre-conditioned so that a conjugate gradient solution converges
rapidly. A numerical demonstration is carried out on the torus and several tests are performed
to validate the solution. We present an alternative method using a kernel-based, multi-scale Laplace
solver to fill the widest gaps followed by a series of conjugate gradient improvement in increasingly
narrow strips around the mask boundary. 
Section 3 discusses a numerical implementation of a preconditioned conjugate gradient method on 
a toroidal domain. Section 4 discusses some of the difficulties related to spherical harmonic
transform on the pixelized sphere particularly near the angular scale of the pixelization.
Section 5 presents an integral equation formulation of the filling in problem in angular space
examining the properties of the kernels first for a toy model and then for more realistic situations.
Section 6 presents an alternative procedure for filling in on the pixelized sphere and Section 7
closes with a few concluding comments.

\section{Constrained Gaussian realizations}

\subsection{Mathematical formulation}

In the previous section we saw that there is considerable interest in 
finding a suitable way to fill in missing data so that full sky filtering 
could be carried out and then masks could be reapplied in order to 
minimize the artifacts of this filling in. Such filling in is inevitably
dangerous as the filled in data is likely to bias the measured non-Gaussianity

In a constrained Gaussian realization, we assume that the power spectrum
of the underlying cosmological model and of the noise is known with sufficient
precision. We generate a random realization according to this distribution 
while constraining the data in the unmasked region to take its observed value. 
Schematically we may write the probability distribution in pixel space
as ${\bf z}=( {\bf x,} {\bf y})^T.$ 
We have 
\ba
P({\bf z}) \propto \exp \left[ -\frac{1}{2} {\bf z}^T({\bf P}+{\bf N})^{-1} {\bf z} \right]
\ea
where ${\bf P}$ is Gaussian primordial power spectrum and ${\bf N}$ is the instrument noise superimposed. 
Setting ${\bf C}^{-1}=({\bf P}+{\bf N})^{-1},$ we obtain
\ba 
P({\bf z}) &\propto &
\exp \left[ -\frac{1}{2} 
\begin{pmatrix} {\bf x}\cr {\bf y}\cr \end{pmatrix} ^T
\begin{pmatrix} 
({\bf C}^{-1})_{\bf xx}& 
({\bf C}^{-1})_{\bf xy}\cr
({\bf C}^{-1})_{\bf yx}& 
({\bf C}^{-1})_{\bf yy}\cr
\end{pmatrix} 
\begin{pmatrix} {\bf x}\cr {\bf y}\cr \end{pmatrix} 
\right] \cr
&\propto &
\exp \left[ -\frac{1}{2} 
\left(
 {\bf x}^T ({\bf C}^{-1})_{\bf xx} {\bf x}
+{\bf x}^T ({\bf C}^{-1})_{\bf xy} {\bf y}
+{\bf y}^T ({\bf C}^{-1})_{\bf yx} {\bf x}
\right)
\right] \cr
&\propto &
\exp \left[ -\frac{1}{2}
\left(
({\bf x}-{\bf x}_{ML})^T
({\bf C}^{-1})_{\bf xx} 
({\bf x}-{\bf x}_{ML})
\right)
\right] 
\ea
where
\ba 
{\bf x}_{ML}=-(({\bf C}^{-1})_{\bf xx})^{-1}({\bf C}^{-1})_{\bf xy} {\bf y}
=-(
{\cal P}_x
{\bf C}^{-1}
{\cal P}_x
)^{-1}(
{\cal P}_x
{\bf C}^{-1}
{\cal P}_y)
{\bf y}
={\bf M}{\bf y}
\ea
where ${\cal P}_x$ and ${\cal P}_y$ are the projection operators
onto the unobserved (masked) and observed (constrained) pixels,
respectively. 

For ${\bf y}$ fixed, the ${\bf x}$ of the constrained realization decomposes as
\ba 
{\bf x}= {\bf x}_{ML}+{\bf x}_R 
\ea
where ${\bf x}_{ML}$ is the deterministic part and ${\bf x}_R$ is the random contribution
of vanishing mean everywhere. Here ML stands for maximum likelihood.

\subsection{Qualitative behavior: an exactly solvable toy model}
\label{Qual:Sect}

Before proceeding to the technical details of a practical and efficient
implementation for generating constrained Gaussian realizations, in order to provide some 
intuition we briefly describe qualitatively what the fields 
${\bf x}_{ML}$
and 
${\bf x}_{R}$ typically look like in the masked region. If we assume 
an exactly scale-invariant spectrum for the primordial contribution ${\bf P}$
(i.e., $P(\ell )\sim \ell ^{-2}$) and no noise (i.e, ($N(\ell )=0,$) then
$C(\ell )=P(\ell )+N(\ell )\sim \ell ^{-2},$ and in this idealized case
${\bf C}^{-1}$ is simply the two-dimensional Laplacian operator $\nabla ^2.$
It follows that in the masked region  $\nabla ^2T=0$ and Dirichlet boundary
conditions are imposed with $T$ fixed on the boundary, which in this idealized
case is of zero breadth, to the values in the unmasked region at the boundary.
For a circular hole of radius $a,$ we obtain the following maximum likelihood (ML) 
extrapolation of the temperature map $T$ into the interior (where $\rho <a$)
\ba 
T(\rho ,\theta )=
\sum _{-\infty}^{+\infty}
\left( \frac{\rho }{a}\right) ^{\vert m\vert }
\int _0^{2\pi }\frac{d\theta '}{2\pi }
\exp \left[ im(\theta -\theta ')\right] ~T(a,\theta '),
\label{Laplace}
\ea
so that the value at the center is the average and the small scale structure on the 
boundary decays as one moves inward toward the center of the hole. When the 
boundary curvature is ignored, solutions have the form
$\exp [ik\theta ]\exp -\vert k\vert (a-r)$ just inside. 

In the presence of white noise (from the instrument), the effective boundary fattens.
When there is white noise, the value measured at the boundary is meaningless because
it is infinitely noisy, and one must average outward a distance of order $\theta _{eq}\approx \ell _{eq}^{-1}$
(where $\ell _{eq}$ is the multipole number at which the white noise and primordial signal are equal). 
In this case, eqn.~(\ref{Laplace}) is replaced with
\ba 
T(\rho ,\theta )=
\sum _{-\infty}^{+\infty}
\left( \frac{\rho }{a}\right) ^{\vert m\vert }
\int _0^{2\pi }\frac{d\theta '}{2\pi }\int _a^\infty d\rho '~
\exp \left[ im(\theta -\theta ')\right] ~K_m(\rho ;\rho ')~T(a,\theta '),
\label{LaplaceB}
\ea
and where $(\rho '-a)/\theta _{eq}\gg 1$ the kernel contributes negligibly. 
As the noise is increased, the support of the kernel spreads outward into
the unmasked region. 

Concerning the random component ${\bf x}_R,$ on small scales and near the center of the hole, the fluctuations
are almost as for an unconstrained realization over both the masked and unmasked domain. But closer to 
the boundary fluctuations are suppressed because some of the fluctuations outside propagate inward
through the evolution of ${\bf x}_{ML}$ described above. Almost all the fluctuations on scales larger
than the dimension of the hole propagate in through ${\bf x}_{ML}.$ In the presence of noise, however,
fluctuations much smaller than $\theta _{eq}$ are only weakly suppressed as one approaches the 
boundary. 

Here we have considered the special case where $P$ is exactly scale invariant. For deviations from
scale invariance one expects the kernel to fan out, even in the absence of noise. The infinitely thin
boundary dependence is an artifact of exact scale invariance. 

\subsection{Efficient computation of random component}
\label{Efficient:Sect}

We would like to generate in an efficient way realizations of the 
random component ${\bf x}_R$ of the constrained Gaussian realization
with zero mean and covariance matrix
\ba 
(({\bf C}^{-1})_{\bf xx})^{-1}
=
{\bf C}_{\bf xx}-{\bf C}_{\bf xy}{{\bf C}_{\bf yy}}^{-1}{\bf C}_{\bf yx}.
\label{starH}
\ea
Generating a random vector with a generic covariance matrix of large dimension
is a costly operation in the generic case. The first step involves finding the 
Cholesky decomposition of the covariance matrix ${\bf C}={\bf GG}^T$ (i.e., the moral equivalent
of taking the square root of a symmetric matrix) where ${\bf G}$ is lower triangular with
strictly positive diagonal entries. This calculation would involve $O(N^3)$ operations
as would matrix inversion. Once the Cholesky decomposition has been calculated, successive
realizations would cost $O(N^2)$ because the bottleneck now is simply a matrix 
multiply. 

This brute force approach can be avoided using a simple trick due to \cite{Hoffman91}. 
If we create a joint random realization of $({\bf x},{\bf y})$ on the full sphere with covariance
matrix 
\ba
{\bf C}=
\begin{pmatrix}
{\bf C}_{\bf xx} & {\bf C}_{\bf xy}\cr
{\bf C}_{\bf yx} & {\bf C}_{\bf yy}\cr 
\end{pmatrix},
\ea 
which is easy because there are no cuts to break the isotropy, then the random variable 
\ba
{\bf x}'={\bf x}-{\bf M}{\bf y}
\ea 
has a covariance matrix equal to 
\ba
{\bf C}_{\bf xx}-{\bf M}{\bf C}_{\bf yx}-{\bf C}_{\bf xy}{\bf M}^T+ {\bf M}{\bf C}_{\bf yy}{\bf M}^T, 
\ea
which for ${\bf M}={\bf C}_{\bf xy}({{\bf C}_{\bf yy}})^{-1}$ yields
${\bf C}_{\bf xx}-{\bf C}_{\bf xy}({\bf C}_{\bf yy})^{-1}{\bf C}_{\bf yx},$
which is 
precisely the desired correlation matrix 
given in eqn.~(\ref{starH}).
Here we have used the block decomposition relation
\ba
\begin{pmatrix}
{\bf A_{xx} }&{\bf A_{xy}} \cr 
{\bf A_{yy}} &{\bf A_{yx}} \cr 
\end{pmatrix} ^{-1}=
\begin{pmatrix}
[ {\bf A}_{\bf xx}- {\bf A}_{\bf xy} ({\bf A_{yy}})^{-1} {\bf A}_{\bf yx} ]^{-1} &
[ {\bf A}_{\bf xx}- {\bf A}_{\bf xy} ({\bf A_{yy}})^{-1} {\bf A}_{\bf yx} ]^{-1} {\bf A}_{\bf xy}({\bf A_{yy}})^{-1}\cr
[ {\bf A}_{\bf yy}- {\bf A}_{\bf yx} ({\bf A_{xx}})^{-1} {\bf A}_{\bf xy} ]^{-1} {\bf A}_{\bf yx}({\bf A_{xx}})^{-1}&
[ {\bf A}_{\bf yy}- {\bf A}_{\bf yx} ({\bf A_{xx}})^{-1} {\bf A}_{\bf xy} ]^{-1} \cr 
\end{pmatrix}
\ea
which holds when ${\bf A}_{\bf xx}$ and ${\bf A}_{\bf yy}$ are invertible,
which is always the case when ${\bf A}$ is positive definite. 
The same relation gives the equality ${\bf C}_{\bf xy}{{\bf C}_{\bf yy}}^{-1}
=( ({\bf C}^{-1})_{\bf xx})^{-1}( ({\bf C}^{-1})_{\bf xy}.$
Note that ${\bf M}$ here is identical to the operator
yielding ${\bf x}_{ML}$ given values ${\bf y}$ in the unmasked
region. Consequently all the difficulty of generating
random realizations reduces to the operation of
finding ${\bf x}_{ML}$ given ${\bf y}.$

\subsection{Conjugate gradient solution and an effective preconditioner}
\label{CG:section}

\newcommand{\norm}[1]{\lVert{#1}\rVert}
The conjugate gradient method, first introduced by 
\citealt{HestenesStiefel} (see also \citealt{GolubVanLoan} for a nice review), 
solves linear equations of the form ${\bf A}{\bf x}={\bf y}$ (where in the original
and simplest case ${\bf A}$ is 
symmetric and positive definite). 
The conjugate gradient method yields a sequence of successive approximations to the exact
solution by forming 
linear combinations of ${\bf y}, {\bf A}{\bf y}, {\bf A}^2{\bf y}, \ldots , {\bf A}^{j-1}{\bf y}.$ The 
condition number, defined as 
$\kappa ({\bf A})=\norm{\bf A}~\norm{{\bf A}^{-1}}$ (where $\norm {\bf T}=\sup _{\bf x\ne 0}\norm {\bf Tx}/\norm{\bf x})$
or the ratio of moduli of the largest to smallest eigenvalues,
plays an important role in determining the rate of convergence of these successive approximations.  
If $\kappa ({\bf A})$ is not too much larger than one, the 
sequence rapidly converges and a good approximation is had after a modest number of 
steps. While the smallness of the condition number is a sufficient
condition for rapid convergence, good convergence properties also
follow if all but a few eigenvalues are 
clustered closely together. 
The fast convergence for small condition
number is easy to understand because in terms of the eigenvalues
one is essentially approximating the real function $f(\lambda )=1/\lambda $ on the 
interval $[1,\kappa ]$ by means of successively higher order polynomials.

For a linear system of dimension $N,$ the conjugate gradient method is guaranteed
to converge after $N$ steps for any value of the condition number. But
this property is not very useful because multiplying by ${\bf A}$ generically
involves $O(N^2)$ operations, so continuing for $N$ iterations
would involve $O(N^3)$ operations, which is the same order as 
a direct method (e.g., Gaussian elimination). On the other hand,
for well-conditioned matrices the speedup resulting from the conjugate gradient method
is considerable, especially in high dimension. When 
the matrix can be applied to a vector using less than $O(N^2)$ operations
(e.g., because of sparsity or some other reason), the speedup is even more
dramatic. 

For our problem, $N\propto n_{side}^2.$ The projection by a mask or its complement entails $O(N)$
operations. Applying the FFT on the rectangular or toroidal region and spherical harmonic 
transform require $O(N\log N)$ and $N^{3/2}\log N$ operations, respectively [\cite{gorski}]. Therefore
solving the problem
\ba
({\cal P}_x {\bf C}^{-1}{\cal P}_x){\bf u}={\bf v} 
\label{ModEq}
\ea
would be ideally suited to the conjugate gradient method if it were not for the fact that
$({\cal P}_x {\bf C}^{-1}{\cal P}_x)$ is extremely ill-conditioned. We estimate the 
condition number $\kappa ({\bf C}).$
Here ${\bf C}={\bf P}+{\bf N}$ where ${\bf P}$ very roughly has a scale invariant spectrum (with $P(\ell )\propto \ell ^{-2}$)
and ${\bf N}$ has a white noise spectrum (with $N(\ell )=({\rm constant})$).
Let $\ell _{eq}$ denote the multipole number where the noise roughly equals the primordial signal. 
It follows that $\kappa ({\bf C}^{-1})$ is approximately $(\ell _{eq}/\ell _{min})^2$ no matter
how fine the pixelization is. The noise serves as a cut-off preventing the small
scale eigenvalues from becoming too extreme. This estimate is not very accurate, but there is no
sense in trying to do better because its value is so extreme (i.e., $\gtorder 10^6$ for Planck)
that the conjugate gradient method on this problem in its present state will not converge fast enough. 

We try to improve the problem by means of a 
pre-conditioner by rewriting eqn.~(\ref{ModEq}) as 
\ba
({\cal P}_x C^{1/2} {\cal P}_x)
({\cal P}_x {\bf C}^{-1}{\cal P}_x)
({\cal P}_x C^{1/2} {\cal P}_x)
{\bf u}'=
({\cal P}_x C^{1/2} {\cal P}_x)
{\bf v}={\bf v}' 
\ea
or
\ba
{\bf M}'~{\bf u}'=({\cal P}_x {\bf C}^{1/2} {\cal P}_x
{\bf C}^{-1}{\cal P}_x {\bf C}^{1/2} {\cal P}_x)
{\bf u}'= {\bf v}'. 
\ea
Here ${\bf v}'$ is easy to obtain from ${\bf v}$ and
so is ${\bf u}$ from ${\bf u}'.$
Since the powers of ${\bf C}$ add up to zero, one would hope
that the eigenvalues of the entire product would cluster near unity,
but the projection operators, which act in real rather than harmonic space, 
could act to spoil this attempt to improve the conditioning of the matrix. 
The sharp boundaries imposed by the projection operator
convert low-$\ell $ modes into a superposition containing some
high-$\ell $ modes (because the initial data to which $C^{-1}$
is applied is taken to vanish on the complement of ${\bf x}).$
Similarly, in the other direction, the projection operator 
also partially converts high-$\ell $ modes into high-$\ell $ modes.
Therefore if mixing were strong enough one could expect 
eigenvalue of order $\lambda _{max}/\lambda _{min}$ and 
$\lambda _{min}/\lambda _{max}$ where $\lambda $ refers to
the eigenvalues of $C$. 
Rather than seeking more sophisticated analytic estimates of the condition number
and convergence rate of this pre-conditioned system, we explore the convergence
by numerical experiment in the next section.

\section{Preconditioned conjugate gradient approach on a periodic flat domain}
\label{PreCondPro}
 
\begin{figure}
\begin{center}
\includegraphics[width=8.5cm]{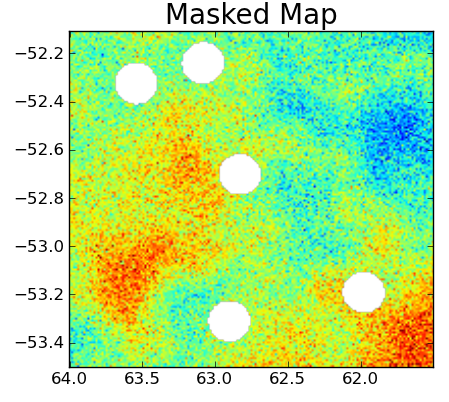}
\includegraphics[width=9.0cm]{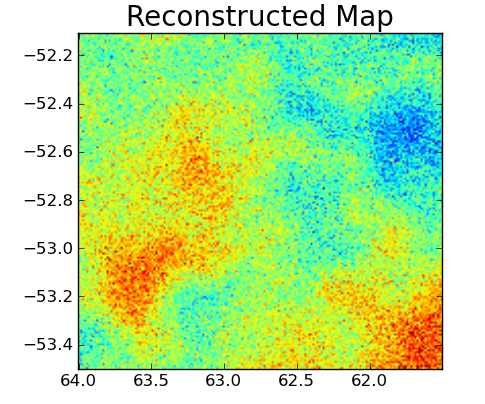}
\includegraphics[width=5.7cm]{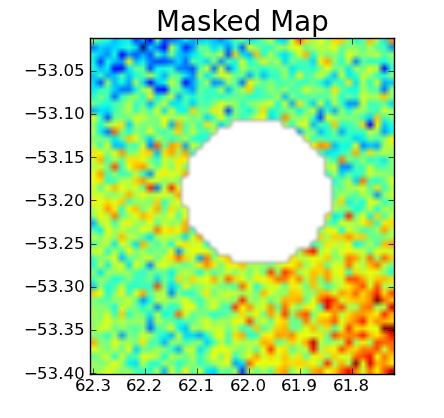}
\includegraphics[width=5.7cm]{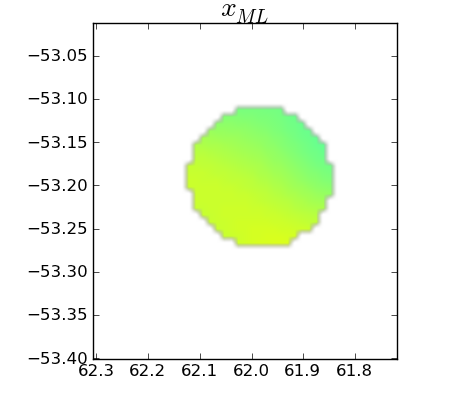}
\includegraphics[width=5.7cm]{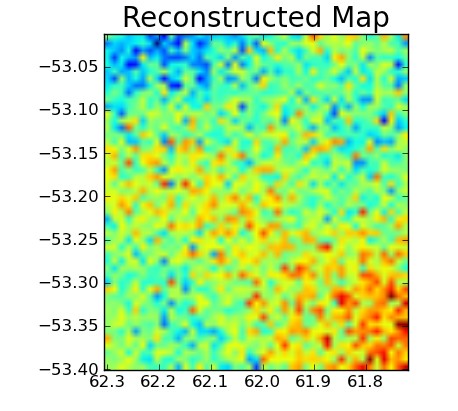}
\includegraphics[width=10cm]{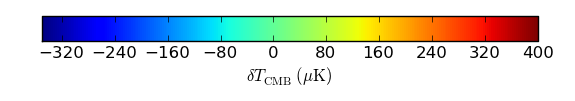}
\end{center}
\caption{
{\bf Conjugate gradient reconstruction of mask regions of mock ACT data.} 
(Top)  The constraints, given by the masked map, and the map reconstructed 
by the mean of Gaussian Constraint realization for a zoom on a 2.1 degree 
square part of the map, the coordinates are given in decimal degree.
(Bottom) we zoom on a typical mask hole, in a region of 0.13 square degree, 
at the center we show the maximum likelihood solution which propagates the information into the hole}
\vskip -10pt 
\label{Fig:CR}
\end{figure}

For high resolution experiments, such as ACT [\cite{Das,Dunkley}] and SPT [\cite{spt}], which maps the small 
scale anisotropies of the CMB on small patches of the sky, it is convenient 
to use the flat sky approximation, where the analysis of the CMB is made on a 
torus. We decide to test our implementation using the characteristics of an 
ACT-like experiment.  The power spectrum of the survey is given by a combination 
of the underlying power spectrum modified by the beam profile coefficients $\omega_{\ell} $
and a white noise term
\ba
P_{\ell}=\omega_{\ell} C_{\ell}+N_{\ell}
\ea 
We simulated 4 CMB maps of $7000\times600$ pixels using this distribution, each of 
them corresponding to a season of observation of 300 square degrees of the sky. 
We applied a realistic point source mask ($\approx$ 400 holes of area $10^{-2}$ 
degree square) to the maps. When we apply a point source mask to a map, we replace the 
pixel values in an area containing each point source with zeros. CMB power spectrum 
estimated from an incomplete sky map is the underlying full-sky power spectrum convolved 
with the power spectrum of the mask. This leads to coupling of modes in the estimated 
power spectrum. The mask becomes even more problematic in the context of lensing 
power spectrum and bispectrum estimation.
We show the result of  one reconstruction in Fig.~\ref{Fig:CR}. 

As expected, the continuity is well preserved, the temperature scale is the same as that
of the initial map and the perimeter of the cut is not visible.
The conjugate gradient algorithm converges extremely quickly, with the residues
following a geometrical progression. For filling in a point source mask, the use of 
a preconditioner is unnecessary. We define the residual for each iteration as 
$ {\bf r}_{(i)}= {\bf y}-{\bf A} {\bf x}_{(i)}$.
We found that the norm of the residual decayed by about a factor of two per iteration.
The reconstruction for one map is then complete in 20 minutes using one processor.


The typical size of the hole is given in angular space by $\ell \approx \pi/\theta$ where $\theta$ 
is the radius of the hole. Here $\ell  \approx 3600$, 
which lies deep in the noise dominated part of the spectrum indicated.
This implies that the maximum likelihood part of the reconstruction mainly continues
the average temperature and its gradient into the hole here. 

%
%
%
%
%
%
%

\begin{figure}
\begin{center}
\includegraphics[width=19cm]{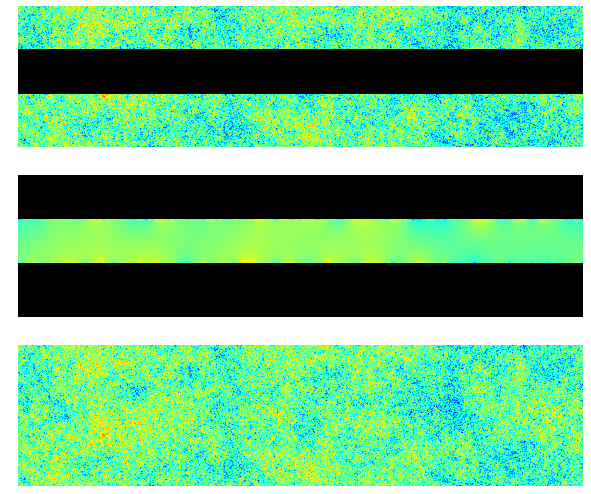}
\end{center}
\caption{
{\bf Reconstruction of Galactic cut}
In the original map (top panel) the pixels in a strip $30^\circ $
wide along the Galactic plane has been masked (shown in black).
The middle panel shows the maximum likelihood reconstruction
of the missing pixels based on the unmasked pixels in the panel above.
In the lower panel show a full reconstruction using a constrained 
Gaussian realization, assuming the underlying power spectrum is known.
}
\label{WideCut1:Fig}
\end{figure} 

\begin{figure}
\begin{center}
\includegraphics[width=14cm]{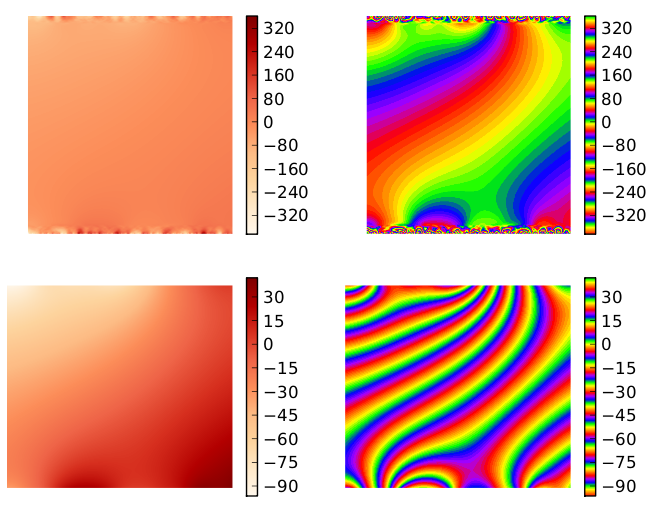}
\includegraphics[width=14cm]{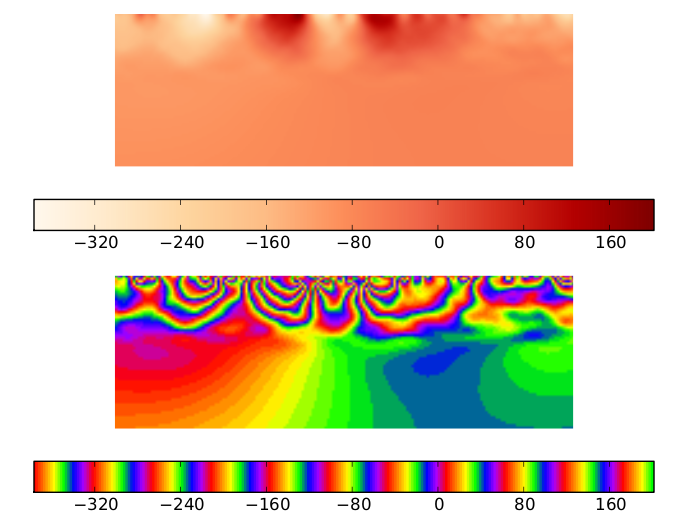}
\end{center}
\caption{{\bf Maximum likelihood reconstruction of the Galactic cut.}
The top row shows the maximum likelihood reconstruction of a square subregion
of the Galactic cut using two color scales, a simple scale on the left
to provide a global view and a periodic scale on the right 
to highlight the contours
of constant value and nature of the variation in the central region.
In the second row the boundary region where the variation is greatest has
been cut away to accentuate the nature of the variation in the central
region, which we call the ``deep interior.'' The bottom row shows a
zoom of the boundary region, where the distinction between the boundary layer and
the deep interior is clearly visible.
}
\label{Fig:Laplace2}
\end{figure}

Fig.~\ref{WideCut1:Fig} shows the reconstruction of a $30^\circ $
wide Galactic type mask, using $N_{x}=4 N_{side}$ pixels in the horizontal direction with $N_{side}=2048$.
The use of a preconditioner for filling in a wide mask is a necessary condition to achieve the convergence in a reasonable amount of time (6 hours of computer time with one processor).
Fig. \ref{Fig:Laplace2}  shows the maximum likelihood reconstruction on  a square subregion of the Galactic cut, enlightening the structural difference between the boundary layer and the deep interior.

The procedure for filling in on the torus described in this section
was applied to the sphere
pixelized using the Healpix package and in particular the forward
and inverse spherical harmonic transforms provided in that package.
With parameters chosen identical to those for the torus, we found
that the iterations of the conjugate gradient solver did not
converge. Although a direct proof is lacking, we strongly
suspect that this failure is due to the lack of Hermiticity of
the operators when approximated using Healpix, under which the harmonic
transform followed by its inverse does not recover the identity
operator exactly. We performed some simulations where the 
standard conjugate gradient algorithm was applied to matrices
with deviations from perfect Hermiticity. We found that 
for modest condition numbers $\kappa \ltorder 100,$ deviations
had to be around 10\%, before the lack of Hermiticity spoiled
or significantly slowed down convergence. However for
larger condition numbers, much smaller departures from Hermiticity were
found to spoil convergence. In this case, the rate of convergence (expressed
as $-\ln (\vert \vert r_{n+1}\vert \vert /\vert \vert r_{n}\vert \vert )$
would decrease with successive iterations and approach zero rather than 
fluctuating around a constant positive value. It could well be that a 
modification of Healpix might remedy this difficulty, but below we describe
another strategy demonstrated to work on the sphere.

We note that for solving $Ax=y$ a number of variations 
have been proposed to the classical conjugate gradient method of 
Hestenes and Stiefel where $A$ is assumed symmetric and postive definite.
When $A^T$ is likewise easily  calculable, one can use the biconjugate
gradient method, where there is no assumption of Hermiticity
[\citealt{fletcher}].
Unfortunately, there is no easy way of applying $A^T$ to
an arbitrary vector. Other Krylov space methods rely on 
calculating a number of inner products of order the square of
the number of iterations, and thus become impractical both because
of storage requirements and because of the 
number of operations required. 
\section{Pixelization errors on the sphere}
\label{PixSphere:Sect}

In this section we explore some of the obstacles to extending the 
conjugate gradient approach that worked so well on the torus to the
sphere. 
As we shall investigate in more detail, many of the nice
properties of the FFT on a rectangular or periodic lattice do not carry
over to the pixelized sphere, and it is these differences that prevent
the constrained Gaussian realizations that worked so well on the torus
from working equally well on the sphere.

On the torus (i.e., a rectangle with periodic boundary conditions imposed)
for a representation of the circle using $N$ equally spaced points,
the discrete Fourier transform has the property that
$$
{{\cal F}_N}^\dagger
{\cal F}_N=I_N
$$
holds exactly and not just approximately. 
Here $I_N$ is the identity operator on the N-dimensional
vector space. 
Mathematically, the fact that
$\{ U(1)\} ^d$ can be approximated by the sequence
$\{ Z_N\} ^d$ as $N\to \infty $ is at the heart of this property.
There thus exists a sequence of arbitrarily fine tilings of
$\{ U(1)\} ^d,$ or equivalently of the $d$-dimensional torus.

The same does not hold for the sphere. Except for the two infinite
series of symmetry groups $C_N$ and $D_N$ (corresponding to the
cyclic and dihedral groups of order $N$ and $2N,$ respectively),
the finite order subgroups of $SO(3)$ are finite in number.
The remaining groups are the tetrahedral (T), octahedral (O),
and icosahedral (I) groups and can be placed in correspondence with
the platonic solids. $T$ is the symmetry group of the tetrahedron,
$O$ of the cube and octahedron, and $I$ of the dodecahedron (12 sides) 
and icosahedron (20 sides). This situation severely limits the regular
tilings of the sphere. The $C$ and $D$ series are not useful because the pixels become
finer only in one dimension.

On the sphere we would like to generalize the 
continuum resolution of unity
\ba
\sum _{\ell =0}^\infty 
\sum _{m=-\ell }^{+\ell }
Y_{\ell m}^*(\hat \Omega )
Y_{\ell m}( \hat \Omega ')
=\delta ^2(\hat \Omega , \hat \Omega ')
\ea
where $\delta ^2(\hat \Omega , \hat \Omega ')=
\delta (\theta -\theta ') \delta (\phi -\phi ') /\sin \theta $,
to something like
\ba 
\sum _{\ell =0}^{\ell _{max}}
\sum _{m=-\ell }^{+\ell }
Y_{\ell m}^*(\hat \Omega _p)
Y_{\ell m}( \hat \Omega _{p'})
=\delta _{p,p'}
\ea
where the indices $p,p'$ label the pixels. 
For the Healpix pixelization, the number of
pixels is given by $n_{pixel}=12\cdot {n_{side}}^2$
where $n_{side}=2^k$ for some integer value of $k.$
The problem is best seen by considering
the discretized orthogonality relation 
\ba 
\sum _p 
{\cal A}_{p}
Y_{\ell 'm'}^*(\hat \Omega _p) Y_{\ell m}( \hat \Omega _p)
=0
\ea
where $(\ell ,m)\ne (\ell ',m'),$
${\cal A}_{p}$ is the area of the pixel $p,$
and $\hat \Omega _p$ is the position of its center
on the celestial sphere.

This relation would be exact in the limit $p\to \infty $ where the discrete
approximation to the continuum integral becomes exact. 
However, it is only possible to obtain 
approximate orthogonality because of the irregularities
in the arrangement of the pixels on the sphere.
Several pixelization schemes for the sphere have been investigated, and a
list of desirable properties may be formulated, but 
unfortunately no single scheme can be found that simultaneously
satisfies all these properties. These properties include:
\begin{enumerate}
\item To the extent possible, all pixels should have the same
shape, which should be as close as possible to a square. 
\item The pixels should all have the same area.
\item The pixels should be aligned along azimuthal circles in order
to allow the FFT to be employed to speed up the 
spherical harmonic transform.
\item Passing to a coarser or finer pixelization should be easy to
calculate and produce sensible results. 
\end{enumerate}
Unfortunately, not all the these requirements can be satisfied, and
certain requirements conflict with others. 

\begin{figure}
\begin{center}
\includegraphics[width=6cm,angle=90]{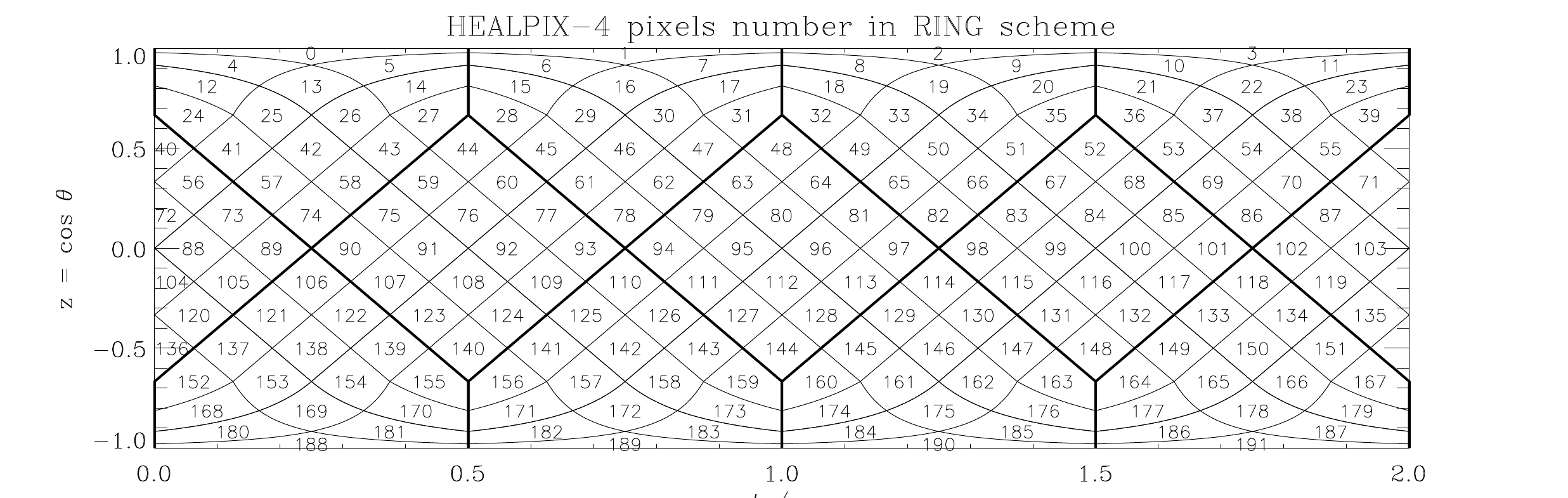}
\end{center}
\caption{
In the Healpix
scheme the sphere is covered with 12 diamond-shaped
regions,
which are in turn subdivided into $n_{side}^2$ pixels where $n_{side}=1,2,4,\ldots .$ 
Here the upper and lower boundaries each represent a single point. 
[Reprinted with permission from Gorski et al. (2005)]
}
\label{Fig:Heal}
\end{figure}

The problems of pixelizing the sphere have been discussed in the literature
by several author and several solutions have been proposed.
\cite{crittenden, crittenden2} proposed an `igloo' pixelization where
the pixels have very nearly the same shape. 
\cite{doroshkevich}, on the other hand, proposed another GLESP scheme.
Presently the most widely used scheme is Healpix, 
\footnote{http://healpix.jpl.nasa.gov}
so we shall restrict
our discussion to this scheme under which 
the sphere is divided into twelve diamonds, which
in turn are subdivided into quadrilateral pixels
in such a way that the pixels all have exactly equal area
and lie on a series of rings of constant latitude, as shown
in Fig.~\ref{Fig:Heal}.
Moreover all pixels on a given ring are equally spaced.
In this way, when the spherical harmonic transform
is carried out, FFTs can be carried out independently
on each ring, requiring an effort of order $n_{side}\ln (n_{side}),$ 
thus greatly speeding up the calculation. The most time-consuming
part is the $\theta $ integral, requiring of order ${n_{side}}^3$
operations. 

Under Healpix the number of pixels per azimuthal ring is
constant over a broad band about the equator, and it is only near
the poles that the number of pixels per ring starts to drop
linearly. 
Consequently, the pixels near the equator are elongated
in the sense along the curves of constant latitude whereas
the pixels at mid-latitude are elongated in the opposite sense---that is,
along the lines of constant longitude.
It is clear that the drop in the number of pixels per ring
at high latitudes leads to aliasing errors. 

On a rectangular lattice the artefacts of the discretization
manifest themselves in a very simple and well-defined way.
All power below the Nyquist frequency is represented perfectly
and any power above this frequency is aliased down into the 
band of representable frequencies. On the sphere, by contrast,
there is no such sharp demarcation. Small wavenumbers are
represented relatively accurately but always with some error.
Very large wavenumbers are completely aliased down, but there
exists a wide grey zone where limited aliasing occurs. The
existence of this intermediate zone is due to the irregular
nature of the arrangement of the pixel centers on the sphere, characterized by
the lack of an equivalent of the reciprocal lattice.

If we fix $n_{side},$ the total number of pixels is
$n_{pix}=12{n_{side}}^2.$  In harmonic space
for the band $0\le \ell \le \ell _{max}$ the total
number of independent real coefficients is
exactly $(\ell _{max}+1)^2.$ Each of the rings in the
equatorial region contains $4n_{side}.$ Since 
representing $\vert m\vert \le \ell $ requires
at least $(2\ell +1)$ degrees of freedom, the 
number of resolution elements per ring is sufficient
only for $\ell $ up to $(2n_{side}-1).$ However,
with this choice the harmonic space includes only 
$4{n_{side}}^2$ degrees of freedom, falling short
of the $12{n_{side}}^2$ by a factor of 3. To match the
number of degrees of freedom between the harmonic
and pixel spaces we would want (retaining only leading terms)
$\ell _{max}=\sqrt{12}n_{side}\approx 3.5 n_{side}.$ 
So one would want approximately $7n_{side}$ pixels
around the equatorial ring whereas one has only
about half that number. 

This factor of two discrepancy is explained by two effects
Firstly the pixel shape varies as a function of latitude
(and also to some extent as a function of azimuthal position). 
At the equator the pixel aspect ratio is 1:1.3 whereas on the 
upper boundary of the equatorial region this ratio is reversed
to 1.3:1 approximately. Here the two numbers indicate 
$(\Delta \theta ):(\Delta \phi ).$ 
Near the polar caps the pixel shape becomes increasingly irregular.
The second effect is
the staggered nature of the pixel center positions between successive
rings. 

This situation is quite different from that of a rectangular lattice,
to which the Nyquist-Shannon sampling theorem can be applied. This
theorem states that a band-limited function can be completely
recovered, with no error, whenever the density of
states implied by the band limit lies below the sampling
frequency (which here is the lattice spacing). The key
perhaps somewhat artificial assumption is that the function
represented is completely band limited, with precisely 
vanishing contribution from higher frequencies. 
When this hypothesis is violated, aliasing occurs. Power from higher
wavenumbers is down converted to power at lower wavenumbers within
the allowed band according to 
$ {\bf k}_{fin}= {\bf k}_{RL}+{\bf k}_{int},$
where the wavevector ${\bf k}_{RL}$ belongs to the reciprocal lattice. 
Since the Shannon-Nyquist theorem applies to one of the two dimensions
of the Healpix pixelization scheme, it can be used to shed some 
light on the pixelization errors that occur when one passes back
and forth between the real space pixelized representation and
a harmonic representation cut-off by a certain 
appropriately chosen $\ell _{max}.$ 

\begin{figure}
\begin{center}
\includegraphics[width=12cm]{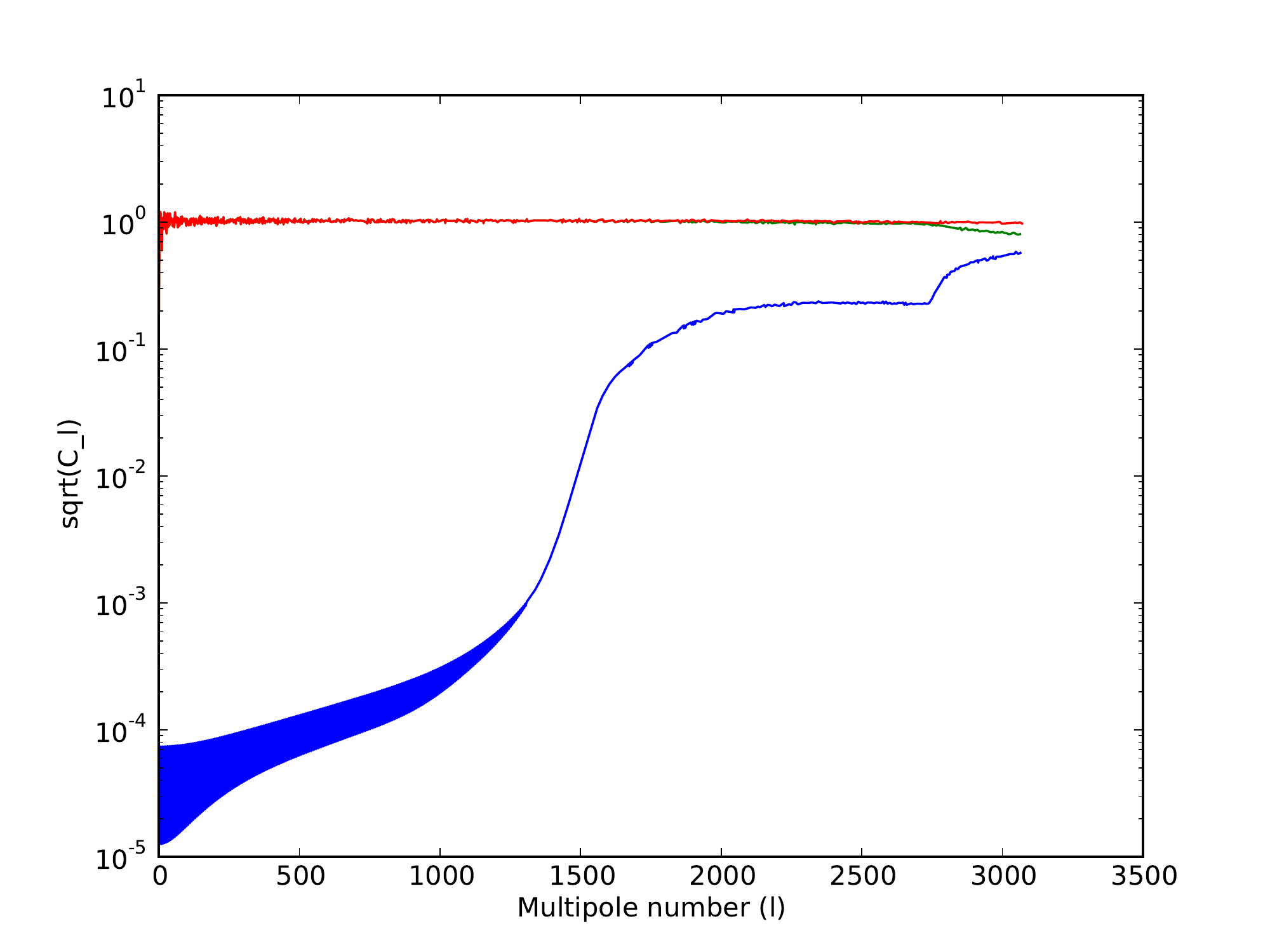}
\end{center}
\caption{
We show quantitatively how the pixelized spherical harmonic transform
followed by its inverse fails at high wavenumber, on scales of order
the interpixel spacing. The red curve represents the input power
spectrum of the map $T.$ After applying the spherical harmonic transform
followed by its inverse, a map $T'$ is obtained, whose power spectrum is
shown in green. The power spectrum of the difference map
$\delta T=T'-T$ is shown in blue. Here we have chosen
$\ell _{max}\approx \sqrt{12}n_{side}$ to match the dimension of the pixel vector space and the
harmonic space. 
}
\label{WN:PS}
\end{figure}

Since representing white noise on the pixelized sphere is an essential
requirement for the calculations in this paper, we illustrate numerically
the extent of the difficulties above by generating a map of
white noise on the pixelized sphere, denoted as $T,$ transforming it to 
harmonic space and back again to obtain $T'(p)=(S_{inv}\circ S\circ T)(p)$
(where $S$ and $S_{inv}$ are the harmonic transform and its inverse, 
respectively), 
and examining the difference $\delta T(p)=T'(p)-T(p).$
For the harmonic transform with $n_{side}=1024,$
we take $\ell _{max}=3546$ to make its matrix
representation as close as possible to a square matrix. 
The map $T'$ has 
81\% of the power of the original map and the difference
map has an rms value per pixel 0.43 (compared to one for the original map),
the value that one would expect if the fluctuations of  
the difference map and of the original map
were uncorrelated. The difference map, however, does display some large-scale
structure at high Galactic latitude aligned with the joints of the 8 polar diamonds
of the Healpix pixelization.
In Fig.~\ref{WN:PS} is shown the power spectrum of the difference map normalized
so that $T(\ell )=1.$ 
We observe that the spherical harmonic transform works well
as long as the spherical function 
is band limited as at wavenumbers well below the limit 
obtained from a simple
counting of the number of degrees of freedom. 

\section{Integral equation formulation in angular space}
\label{IntegralKernel:Sect}

\begin{figure}
\begin{center}
\includegraphics[width=8cm]{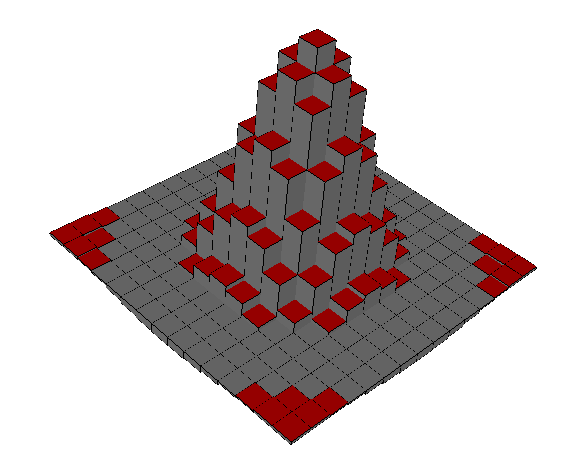}
\includegraphics[width=8cm]{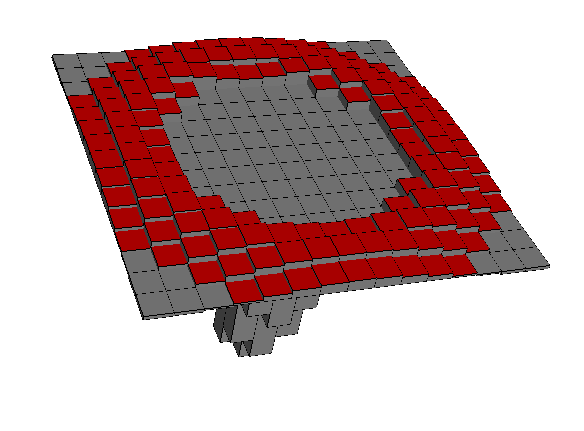}
\end{center}
\caption{
{\bf Maximum likelihood kernel.}
We plot the kernel ${\cal K}(r)$ on a grid slightly smaller than 
$0.5^\circ $ squared. Each box represents a square pixel 
of the area $(1.72 ~ {\rm arcmin})$ corresponding to the area
of an $n_{side}=2048$ pixel. Here we have assumed 
an experiment with a 5 arcmin fwhm resolution and a noise level
of $15\mu K\cdot $arcmin, which is slightly better than Planck.
More noise and a fatter beam would both widen this kernel.
A view from above and below highlight the slight oscillations
in the kernel, which has a radius of about 5'. 
}
\label{Fig:Kern}
\end {figure}

In Sect.~\ref{Qual:Sect} we discussed the qualitative behavior of the 
maximum likelihood solution within the masked region under the 
assumption of an exactly scale invariant power spectrum and no instrument
noise. In this section we generalize first to the case of 
an exactly scale-invariant spectrum with white noise added, and
then to the general case where the power spectrum deviates
from scale invariance in an arbitrary way. Whereas in the case 
of exact scale invariance without noise one would obtain an 
integral equation formulation with a singular kernel, the presence
of noise provides the kernel with a nonzero width and deviations from scale invariance
provide other interesting structure. We present an integral equation
formulation of the problem in this section and show that the 
kernel is short-range even though there is a falling off tail to
arbitrary distance. 
The special case of an exactly scale-invariant spectrum with no noise could be 
described as the limit of Gaussian kernels whose width tends to zero. 

In order to obtain exact analytic expressions,
we assume for the moment an exactly scale-invariant power spectrum
for the primordial contribution with perfect white instrument noise,
so that
\ba 
\left( \frac{1}{P+N}\right)
(\ell )=
\left( 
\frac{A}{\ell ^2}+\frac{A}{\bar \ell ^2}
\right) ^{-1}
=\frac{\bar \ell ^2}{A}
\left[ 1-\frac{\bar \ell ^2}{\ell ^2+\bar \ell ^2}\right] .
\label{RSeqn}
\ea
Here $\bar \ell $ indicates the multipole number where the noise and the primordial signal are equal.

\begin{figure}
\begin{center}
\includegraphics[width=8cm]{%
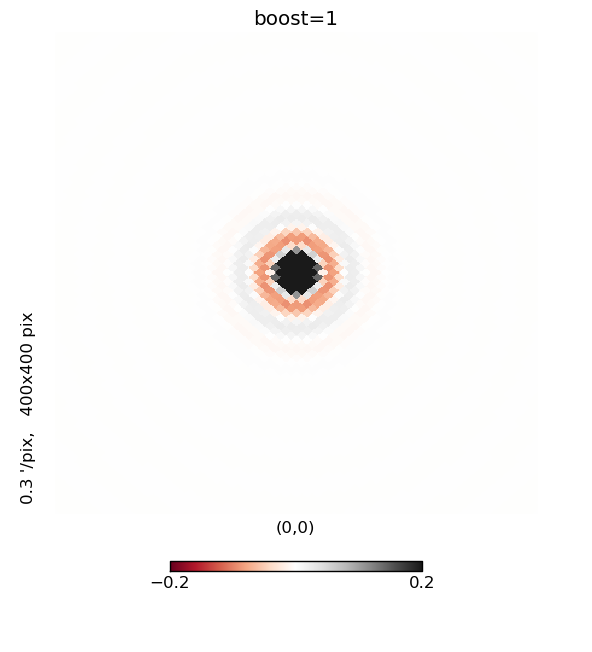}
\includegraphics[width=8cm]{%
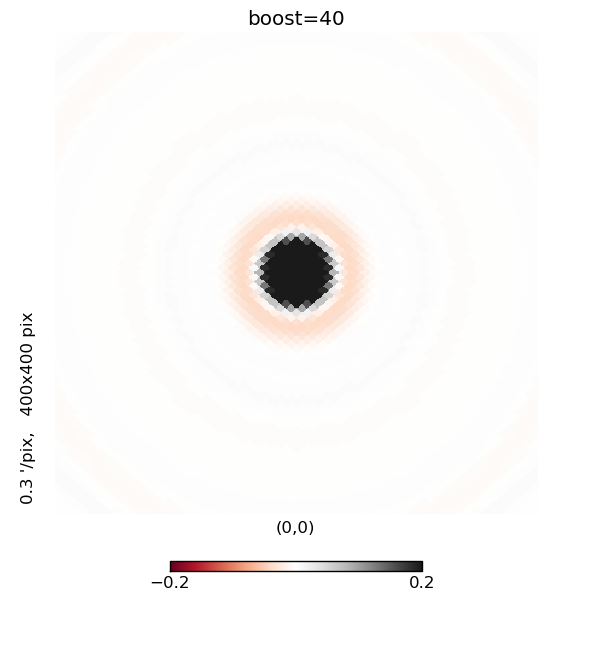}\\
\includegraphics[width=8cm]{%
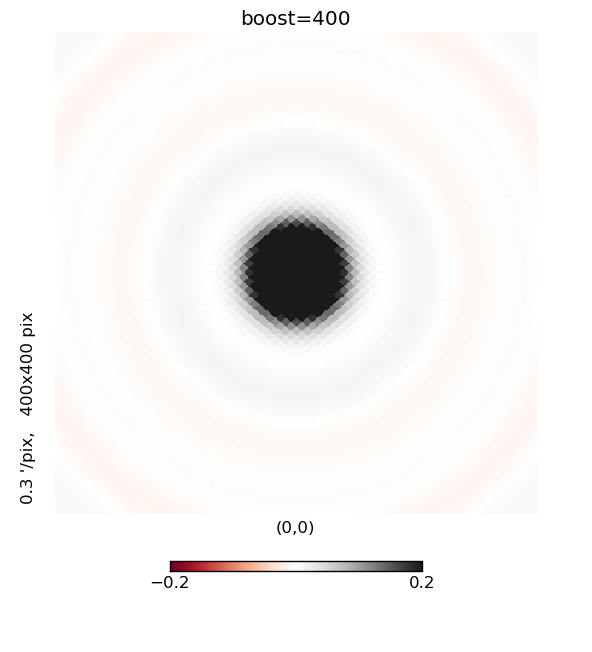}
\includegraphics[width=8cm]{%
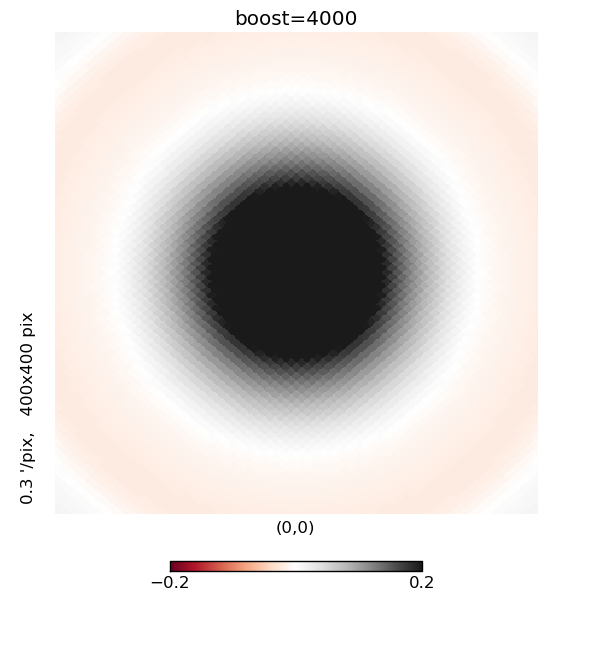}
\end{center}
\caption{{\bf Kernels for maximum likelihood filling in.}
From left to right and then top to bottom we show the
kernels with the instrument noise boosted by
factors of $1,$ $40,$ $400,$ and $4000,$ respectively.
The square patches are 120 arcmin on a side.
A color scale has been chosen so that white corresponds
to a vanishing value, redish hues to negative values, and greyscale
hues to
postive values. The kernel maximum has been normalized
to +1 and the color scale saturates at $\pm 0.2$ in order to
emphasize the oscillations surrounding the central peak.
The size of the central peak, which has no interesting structure,
is exaggerated by saturation effects.
The individual pixels resulting
from an $n_{side}=2048$
pixelization are clearly visible.
}
\label{Kernels:Fig}
\end{figure}

\begin{figure}
\begin{center}

\includegraphics[width=7cm]{%
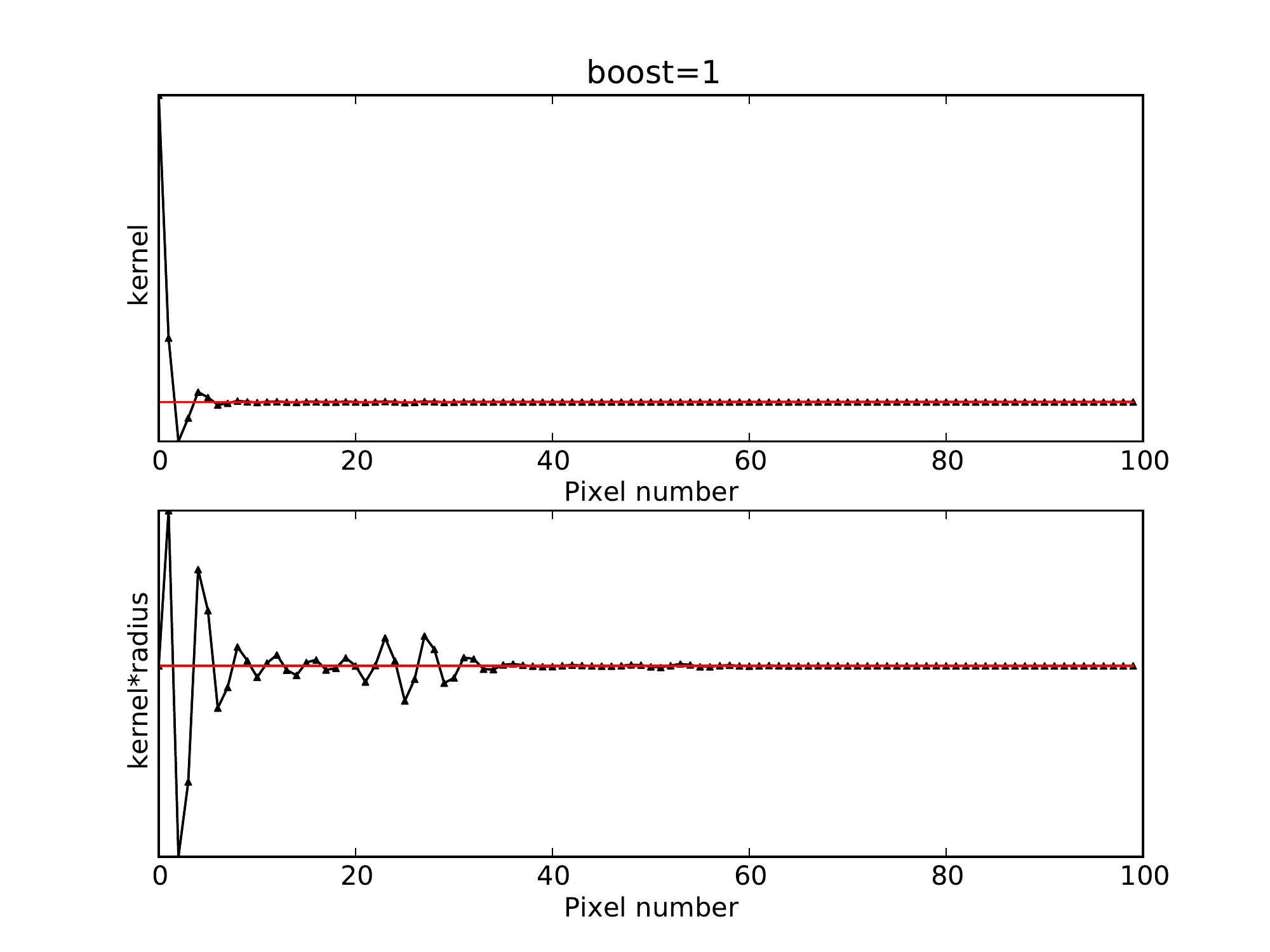}
\includegraphics[width=7cm]{%
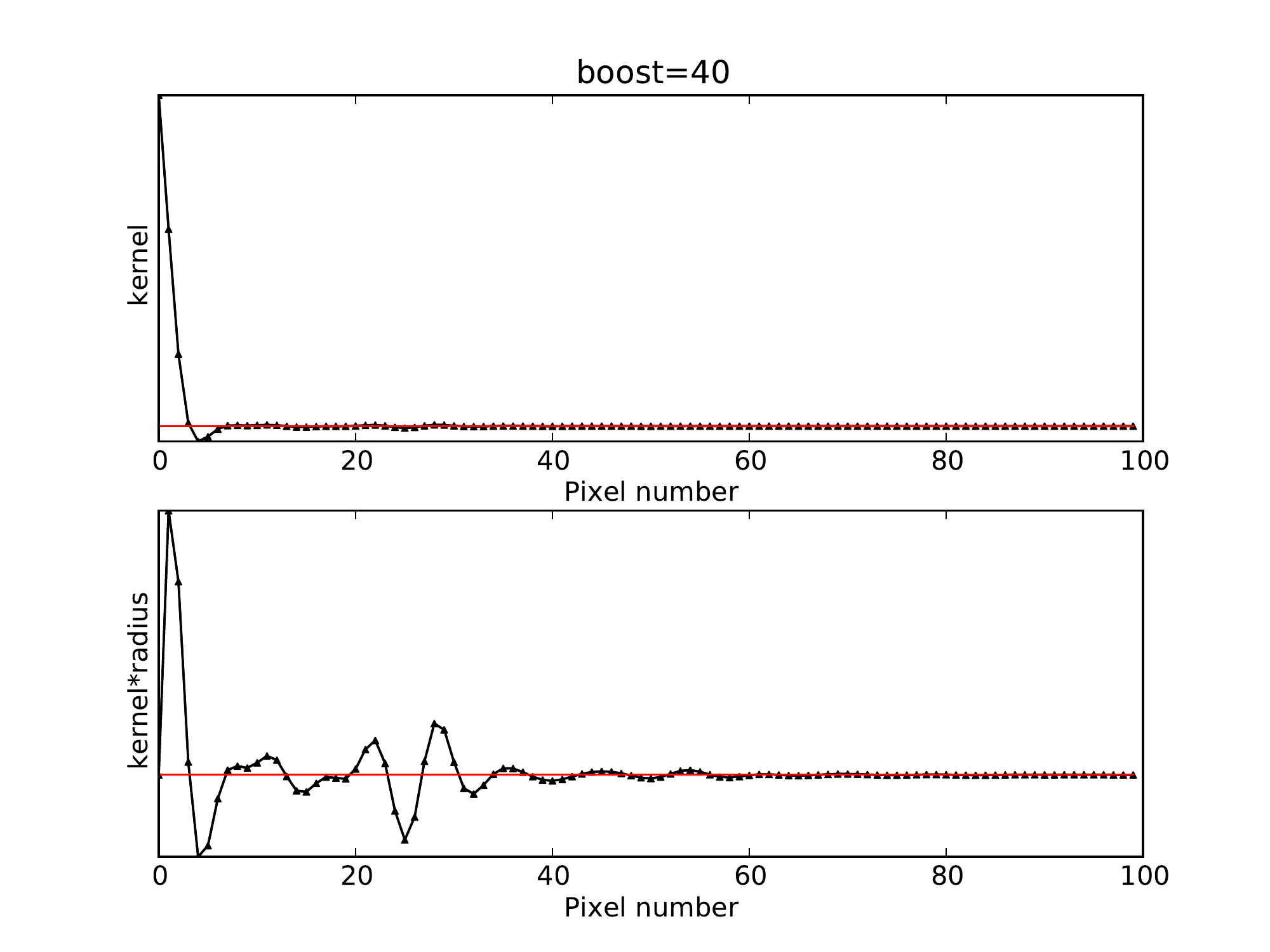}\\
\includegraphics[width=7cm]{%
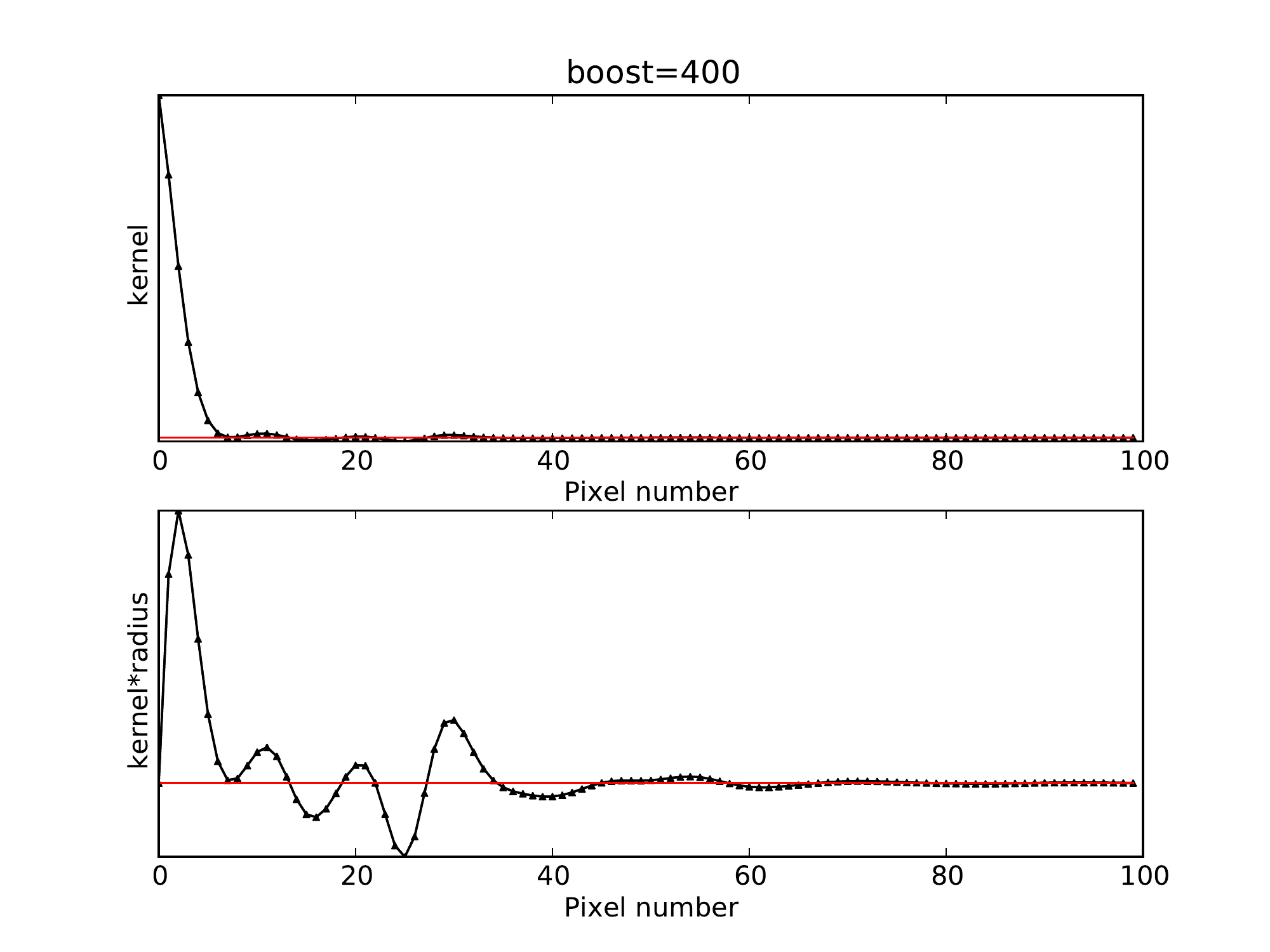}
\includegraphics[width=7cm]{%
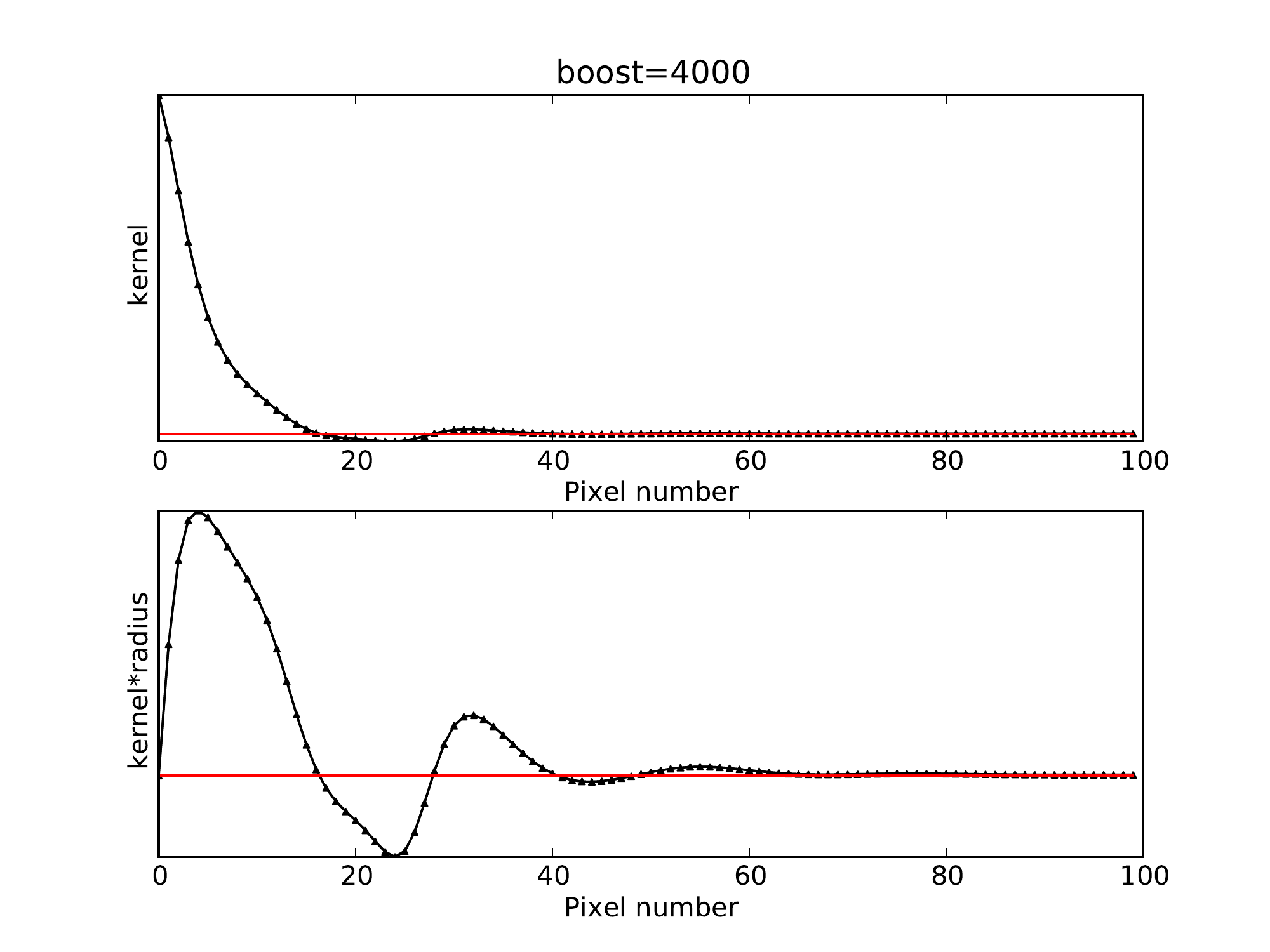}
\end{center}
\caption{{\bf Kernels for maximum likelihood filling in.}
For the same kernels as in the previous plot we show
the radial profiles, first plotted as $K(r)$ and then
as $r\cdot K(r)$ to highlight the radial oscillations
and to give a weighting that includes the effect of the volume element.
For the kernel with unboosted noise the oscillations have an important
effect whereas for the boosted kernels the effect of
the oscillations is less important.
}
\label{Kernels:Fig2}
\end{figure}

\begin{figure}
\begin{center}
\includegraphics[width=12cm]{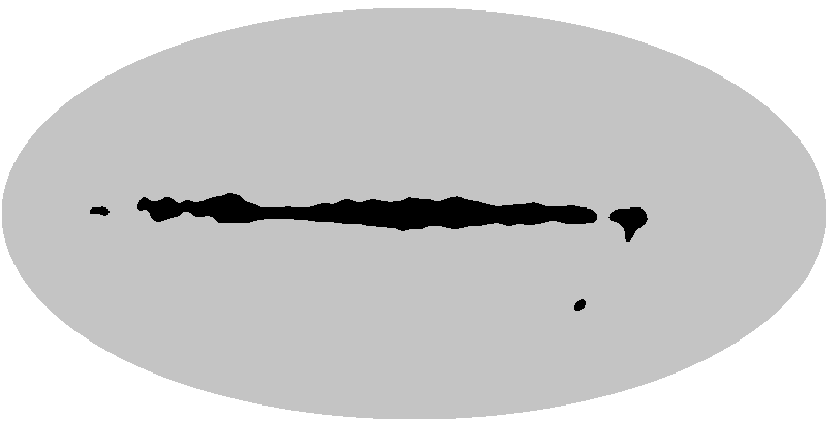}\\
(a)\\
\includegraphics[width=12cm]{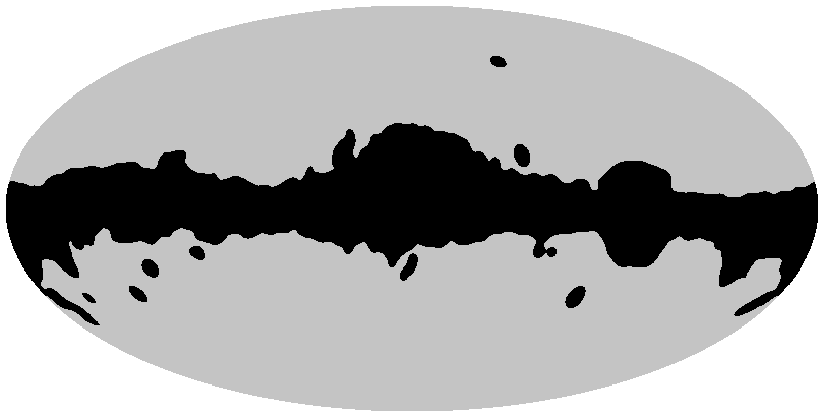}\\
(b)\\
\includegraphics[width=12cm]{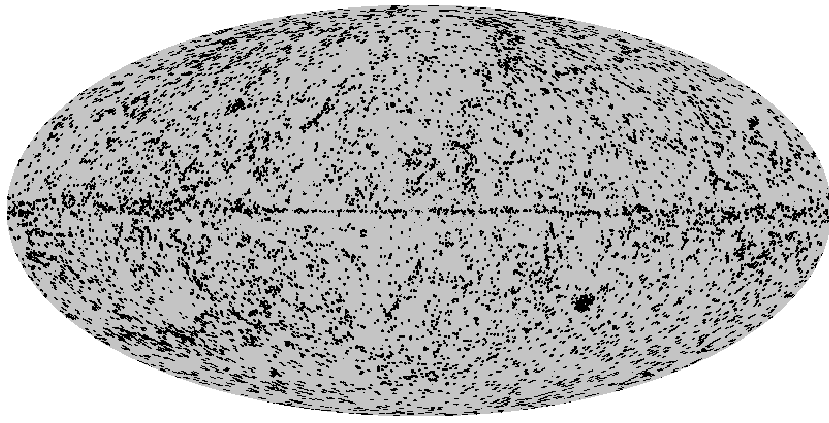}\\
(c)
\end{center}
\caption{{\bf Three candidate Planck masks.} 
In panels (a) and (b) the Galactic plane is masked out less and 
more aggressively and in panel (c) a mask for the discrete sources of the Planck ESRC is shown.
{\it (Courtesy of Jacques Delabrouille and Guillaume Castex)}}
\label{mask:Fig}
\end{figure}

For a realistic power spectrum the kernels would be slightly different
and would have to be calculated numerically.
It is convenient to assume a continuous random
field that has not been pixelized. Here the first term represents the 
singular part (here a $\delta $ function) and the second term the smooth regular
part. Consequently the probability of a configuration or map $T(z)$
is proportional to
\ba
P\Bigl( T({\bf z})\Bigr)\propto 
\exp \left[
-\frac{\bar \ell ^2}{2A}\int d^2z\int d^2z' T(z)
\left\{
\delta ^2(z-z')-{\cal K}(\vert z-z'\vert )
\right\}
T(z')
\right]
\label{kern:rel}
\ea
where 
\ba 
{\cal K}(r)=\bar \ell ^2\int \frac{d^2\ell }{(2\pi )^2}
\frac{\exp [i\boldsymbol{\ell }\cdot {\bf r}]}{\ell ^2+\bar \ell ^2}
=\frac{\bar \ell ^2}{(2\pi )}\int _0^\infty \frac{t~dt~J_0(\bar lrt)}{t^2+1}
=\frac{\bar \ell ^2}{2\pi }K_0(\bar \ell r).
\ea 
Since 
\ba
(2\pi )\int _0^\infty r~dr ~{\cal K}(r)=1,
\ea
the convolution with ${\cal K}$ is simply a weighted average over neighboring 
pixels. 
We recall the following asymptotic properties of the modified Bessel function
$K_0(w):$ 
$K_0(w)\approx -\ln [w/2]-\gamma $ for $0<w\ll 1$ and 
$K_0(w)=\sqrt{\pi /2w}\exp [-w]$ for $w\gg 1$ where $\gamma $ is Catalan's constant.
Note that the logarithmic divergence is very mild and dwarfs the factor $r$
in the area element, so the bulk of the weight of this average arises from
pixels a distance of order $\bar r=\bar \ell ^{-1}$ from the center.
The kernel dies off as $\exp [-r/\bar r],$ so filling in is 
an extremely local process in this special case. We will see that
for more realistic power spectra (e.g., with acoustic peaks) the 
fall-off is less rapid. 

It is convenient to split ${\bf z}$ into masked and unmasked 
(or equivalently unobserved and observed )
pixels ${\bf z}=({\bf x},{\bf y}).$ As discussed earlier, 
the bottleneck in carrying constrained Gaussian 
realization is finding ${\bf x}_{ML}$ given a set of values for ${\bf y}.$
We want to find ${\bf x}_{ML}$ given values for ${\bf y}$ so that
the integral in the exponential of eqn.~(\ref{kern:rel}) is minimized. A 
necessary and sufficient condition is that for all ${\bf x}$
\ba 
T({\bf x})=\int d^2z~{\cal K}(\vert {\bf z}-{\bf x}\vert )T({\bf z})
\label{Boo}
\ea
where the integral over ${\bf z}$ includes masked and unmasked pixels alike.
In Fig.~\ref{Fig:Kern} the numerically computed
kernel for the Planck experiments is shown
assuming the instrument specification given in the caption and the 7-year 
WMAP best-fit cosmological model rather than the idealized assumptions
above. Figs.~\ref{Kernels:Fig} and \ref{Kernels:Fig2} show the 
shapes of the kernels for the Planck noise level and also several
boosted noise levels. 

The short-range character of the kernel makes manifest that filling in is
a relatively local process, despite the fact that low-$\ell $ multipoles
contribute substantially to the values inside the widely separated holes.
Widely separated holes can be filled in independently with very little
error because their coupling decays as $\exp [-d/\bar r].$
The problem of coupling may in principle be solved iteratively by generating a 
sequence of maps 
$T^{(i)}=( T^{(i)}_x, T_y )$ where
\ba 
T^{(i+1)}_x({\bf x})=
\int d^2z ~ {\cal K}(\vert {\bf z}-{\bf x}\vert )~
T^{(i)}(z).
\label{Algo}
\ea
This is akin to Gauss-Seidel iterative relaxation for elliptic PDEs
approximated using a finite-difference scheme.

Now we pass to some orders of magnitude first assuming a single isolated
hole of dimension $D.$
The convergence of the algorithm in eqn.~(\ref{Algo}) would be controlled by
the dimensionless parameter $\lambda =(D/\bar r).$
The convergence is roughly geometric after an initial transient, and 
the number of steps required for convergence scales as $O(\lambda ^2).$
For holes of oblong dimensions (e.g., an elongated rectangle or even 
an infinite strip), it is the smaller dimension that should be used
for $D$ in the expression for $\lambda .$ The convergence scales as $\lambda ^2$ because settling
down is essentially a diffusive process, where the random step per 
iteration is $\bar r.$ 

Eqn.~(\ref{Boo}) is remarkably similar to a relation for solutions to Laplace's
equation in two dimensions, where 
\ba 
\phi (x,y)=
\int _0^{2\pi }\frac{d\theta }{2\pi }\phi (x+\rho \cos \theta ,y+\rho \sin \theta )
\ea 
whenever the disk of radius $\rho $ about $(x,y)$ lies entirely within the 
domain where Laplace's equation is satisfied. 
For a harmonic function, because of the above relation and the symmetry of the kernel,
eqn.~(\ref{Boo}) would automatically hold.

This analogy suggests that the farther one goes into the interior of 
a masked region, staying several lengths $\bar r$ from the boundary, 
the maximum likelihood reconstructions looks increasingly like a solution
to the Laplace equation, for which any kernel is sufficient to reconstruct
the interior. This feature suggests a method of accelerating the 
convergence of eqn.~(\ref{Algo}) with very little error. 
The idea is to use in the 
far interior kernels fatter than ${\cal K}$
to accelerate the propagation, or diffusion, of information from the boundary region
into the interior. It is only within the boundary layer (within a few lengths
$\bar r$ from the unmasked region) that we expect significant deviations from Laplace type
behavior, because the kernel looks beyond the boundary.

\section{A practical filling in procedure for full-sky spherical maps}

Fig.~\ref{mask:Fig} shows three representative masks of the sort likely to be used
for the Planck analysis. From top to bottom are shown:
(a) a mask cutting out a modest portion
of the galaxy, (b) a more aggressive Galactic mask, and (c) a point source mask. In order to
define the filling in problem, we must also set a level for the instrument noise. The 
filling in problem is numerically more demanding for smaller values of the instrument noise. 
Therefore we adopt a beamwidth of 5' (fwhm)
for all the channels and combine the 100 GHz, 143 GHz, and 217 GHz channel sensitivities
in quadrature and assume a 30 month survey
to obtain a white noise level of $15\mu K\cdot {\rm arcmin}$ in order to formulate a test
problem slightly more difficult than what will be encountered by Planck. 
As discussed previously, the most challenging mask is (b) with the wide Galactic cut. 
Since the width of the kernel is of order 6' and that of the cut is about $40^\circ ,$ 
$\bar \lambda =400$ and the number of iterations required for converge would be of order
$\textrm{(few)}\times 10^6$ using the iterative rule in eqn.~\ref{Algo}. This
is clearly impractical except for perhaps narrow point source masks. Therefore 
an approach based on Gauss-Seidel iteration is not feasible.

In this section we describe a two-part procedure. First the far interior of
the large holes is filled using a Laplace solver 
especially suited to domains of the pixelized sphere with irregular boundaries.
We already showed in
Sect.~\ref{IntegralKernel:Sect} why the Laplace equation holds asymptotically as one passes 
into the far interior of the masked region. In the next subsection we investigate this question quantitatively to
determine exactly how far into the interior one must go for this to be an accurate
approximation. Then we describe the details of our fast Laplace solver for the pixelized sphere.
The Laplace approximation is good as long as the scales over which the maximum likelihood
solution varies is large compared to the width of the exact kernel. As one passes farther
into the interior, the small scale structure on the boundaries decays
away and this becomes a better and better approximation. But as one approaches the boundaries
of the masked regions, the length scale of the fluctuations becomes smaller and smaller
almost in a fractal way (until the noise acts as a cutoff). The central region of the kernel
primarily describes the effect of the noise, and the oscillations in the tail where the kernel takes
negative values reflect the acoustic oscillation structure. 
That this qualitative description is correct is clearly illustrated by the preconditioned 
conjugate gradient solutions on a flat domain discussed in Sect. \ref{PreCondPro} and shown in 
Fig.~\ref {Fig:Laplace2}. (See in particular the zoom of the mask boundary.)
See also Fig. \ref{FillInHealFig}.

The second step of our
procedure corrects the approximate solution near the boundary by using a kernel to improve
the small-scale structure between the far interior region and the unmasked region. 
The kernel captures those aspects of the maximum likelihood not well approximated by
the Laplace approximation.
If these strips
are too wide, the condition number $\kappa $ of the resulting linear system is too large, and 
we cannot achieve convergence within a feasible number of iterations. 
We use several strips of decreasing width and 
artificially boost the noise to reduce the condition number on all but the last and narrowest strip. In this
way information about the power spectrum on larger scales can be taken into account
but the small scale structure is washed out. In the later strips
closer to the boundary with less noise 
the correct small scale structure is restored.
Fig.~\ref{FillInHealFig} illustrates this noise boosted kernel procedure on a strip $2.9^\circ $ wide 
on the celestial sphere.

\subsection{Accuracy of Laplacian approximation in far interior of masked region}

In Section \ref{IntegralKernel:Sect} drawing inspiration from an exact analytic 
expression for the kernel, we argued that in the far interior
of the masked region---that is, a few times $\bar r $ away from the boundary---
the maximum likelihood solution should be well approximated by a solution 
to the Laplace equation. To test this assertion we must establish an appropriate figure
of merit for evaluating this approximation. 
The most appropriate test is to propagate boundary 
values into the interior of a region lying completely within the far interior
of the large hole assuming Laplace's equation
and compare to a reliable numerical solution calculated
using the conjugate gradient method.  

The Laplace interpolation from the boundary of a periodic
rectangular strip can be calculated in a very simple way.
Suppose that we want to solve the Laplace equation on a cylinder 
with $0\le x\le a$ where $x=0$ and $x=a$ are identified
and $0\le y\le b.$ The solution may be continued from the two
boundaries into the interior using
\ba
\phi (x,y)=\sum _{m=-\infty }^{+\infty }
\int _0^a \frac{dx'}{a} 
\exp [im(2\pi /a)(x-x')]\left[
\frac{\sinh [(2\pi m/a)(b-y)]}{\sinh [(2\pi m/a)b]}
\phi (x',y=0)
+
\frac{\sinh [(2\pi m/a)y]}{\sinh [(2\pi m/a)b]}
\phi (x',y=b)
\right] .
\ea
For a pixelized surface with $N$ points in the $x$
direction, the sum becomes restricted from $0$ to $(N-1).$

\begin{figure}
\begin{center}
\includegraphics[width=18cm]{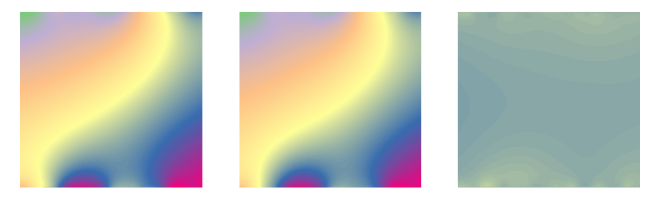}\\
\includegraphics[width=10cm]{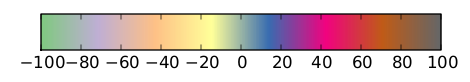}
\end{center}
\caption{{\bf Laplace's equation as an approximation for the far interior of the hole.}
We demonstrate how well Laplace's equation is satisfied in the far interior of a large
masked region, here the $40^\circ $ cut, by cutting away the boundary region $2^\circ $
wide, as shown in the left panel, and using the boundary data here on top and below combined
with Laplace's equation to calculate an interior solution, shown in the middle panel.
The right panel shows the difference, which is small ($<5\% $).
}
\label{Fig:Laplace8}
\end{figure}

In 
Fig.~\ref{Fig:Laplace2}
we showed a setup on the torus designed to mimic the problem of filling in
an aggressive Galactic cut. 
In Fig.~\ref{Fig:Laplace8} we show how the exact solution compares with the
Laplace solution on the domain resulting after $2^\circ $ along each of 
the boundaries has been removed from the originally $20^\circ $ wide
strip. The left, middle, and right panels show the exact conjugate
gradient solution, the Laplace solution, and the difference map,
respectively. We observe that the error is very small. 
We see that the Laplace equation is 
a very good approximation in the far interior.

\subsection{Automatic domain definition in the masked region}
\label{DefineMask:Sect}

\begin{figure}
\begin{center}
\includegraphics[width=19cm]{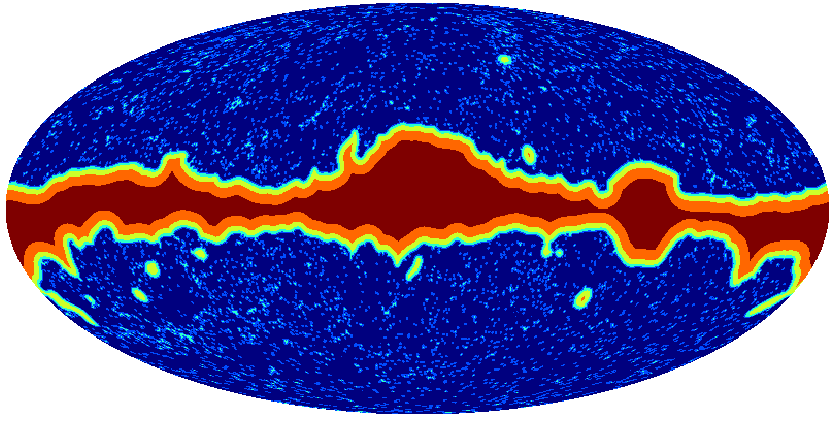}
\end{center}
\caption{{\bf Subdivision of masked sky according to distance from unmasked region.}
For the kernel based methods of solving the Laplace equation in the masked region using the
data in the unmasked region as boundary data and for the improvement of the maximum likelihood
solution near the boundary using the exact kernel and kernels with boosted noise, it is 
essential to subdivide the masked region into subregions according to distance from the unmasked 
region. Here is shown the result of the algorithm described in the text. The dark blue is the
unmasked region and the regions going from light blue to redish brown are subregions farther and
farther from the boundary.
}
\label{Subdivision:Fig}
\end{figure}

One of the required tasks, both for the Laplace solver and the 
noise-boosted kernel improvement of the solution near the boundary,
is to subdivide the masked region into subdomains consisting of
points closer than a certain distance $\delta $ from the boundary. 
Let us call the unmasked, visible region $R_0.$
We define a boundary layer $R_1$ consisting of those pixels within
a distance $\delta _1$ of $R_0.$ 
Masks to avoid the galaxy and point sources 
are often quite irregular and complicated in shape, so
a robust automatic procedure is needed for subdividing the mask.
We calculate $R_1$ by setting all the pixels in $R_0$ equal to
one and all the pixels in its complement equal to zero. Then we convolve
this map 
with a top hat kernel of radius $\delta _1/2,$ and all the pixels with a positive
value and not in $R_0$ belong to $R_1.$ Those farther in the interior should be 
still null after the convolution. In practice a small positive threshold 
(which we choose equal to 0.03) is used to avoid numerical artifacts. 
This procedure is repeated to partiton points in the masked domain into
a number of subdomains according to their closest distance from the boundary. 
The result of the subdivision of the masked region
for the combined Galactic and point source masks in Fig.~\ref{mask:Fig} 
is shown in Fig.~\ref{Subdivision:Fig}.  

\begin{figure}
\begin{center}
\includegraphics[width=6cm]{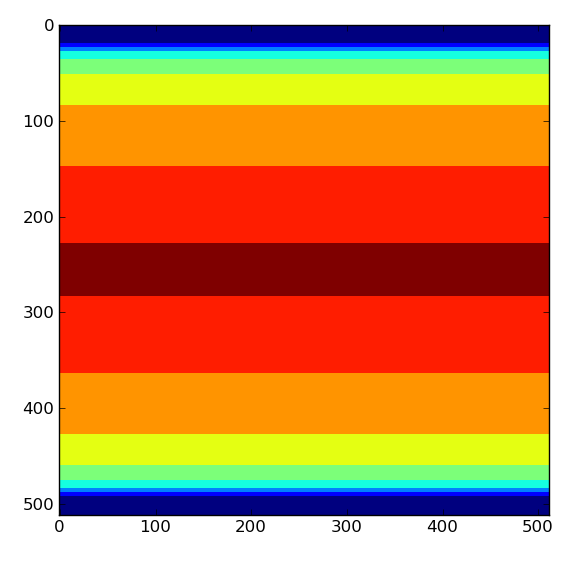}
\includegraphics[width=6cm]{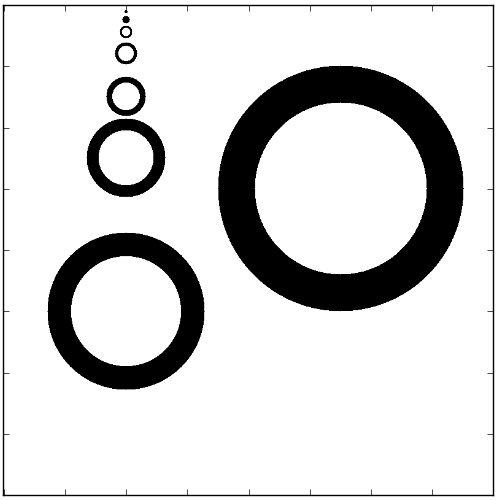}
\includegraphics[width=4cm]{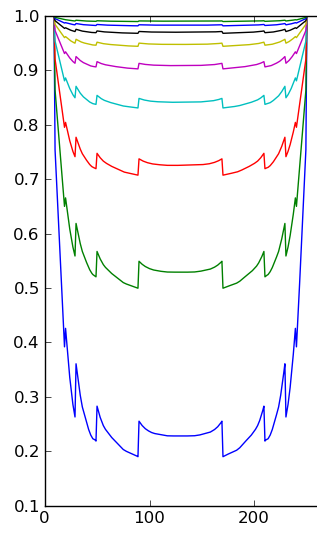}
\end{center}
\caption{{\bf Laplace solver scheme.}
The Laplace solver described in the text uses 9 kernels of various widths 
(shown in the middle panel).
The left panel shows the unmasked region (in dark blue) and several nested subregions of increasing
distance from the boundary. Only in the first very thin subregion is the kernel allowed to extend
into the unmasked region (in order to transmit the boundary data in the edge of the masked region).
In all but the outermost of the masked subregions the largest kernel that does not extend into the 
unmasked region is used to update the interior data by replacing a value with the average over the
kernel. The rightmost panel illustrate the convergence with successive iterations. The jaggedness
arises from the discontinuous change in the choice of
kernel as one passes from one subregion into the neighboring
subregion. Here the boundary data was set to one and the interior values to zero as an initial
condition. Numerous other tests with varying wavenumber for the boundary data were carried out
to validate the accuracy of the method and characterize its convergence properties.
}
\label{LaplaceSolver:Fig}
\end{figure}

\begin{figure}
\begin{center}
\includegraphics[width=18cm]{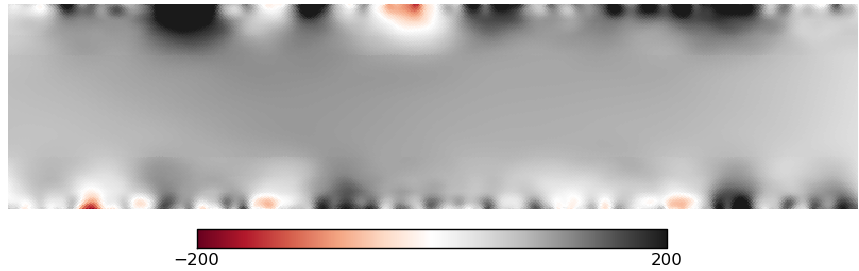}
\includegraphics[width=18cm]{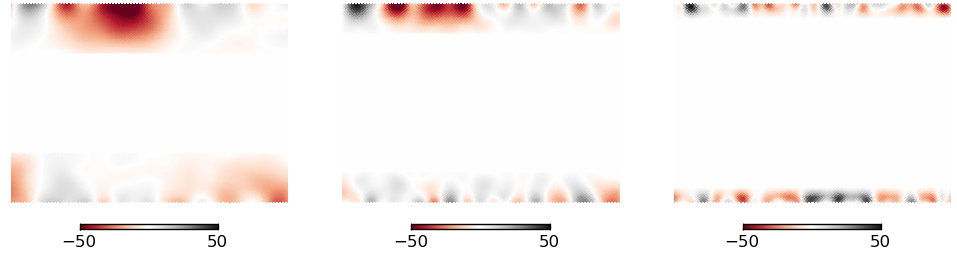}
\end{center}
\caption{{\bf Filling in on the sphere.} We show the maximum likelihood
filling in of $2.9^\circ $ ($\approx $100 pixel) wide equatorial strip 
of the celestial sphere pixelized using Healpix at $n_{side}=2048$ 
resolution using the parameters described in the text. First the 
filling in is carried out assuming a noise level boosted by 2000 in 
order to fill in the large scale structure in the middle of the strip. 
Then the regions near the boundaries are corrected to fill in the small
scale structure by a sequence of three refinements across narrow strips
just above and below the upper and lower boundaries. The widths
and noise boosts of these strips are $(\beta _{noise}, w_{pixels})=$ 
$(400,35),$ $(40,15)$ and $(1,7).$ The bottom three panels show
the difference maps between successive iterations (left) to (right).
}
\label{FillInHealFig}
\end{figure}

\subsection{Laplace solver}

There are many ways to solve the Laplace equation on a discretized domain (e.g., multigrid, finite-element
methods, etc.), but for the present problem the choice of method is dictated
by the irregular nature of the boundary. A method is needed that will work with 
a masked defined as a boolean function defined on a pixelized sky. The method
described below closely resembles multi-grid but is kernel based so that it can
easily be implemented on the pixelized sky. The most naive method---relaxation across
neighboring pixels---would require of order $w^2$ iterations where $w$ is the width
of the domain measured in pixels. We require a method that converges in a number 
of iterations of order a power of $\log (w).$ This rapid convergence is achieved 
by employing a series of kernels of varying radius. We depart from the observation
that the value of a harmonic function in a circle is equal to the average value 
of the function on the circumference. Consequently, in solving by relaxation one 
wants to update values in the interior by using the largest circle that does not
extend outside the masked domain. We also need to propagate information from the
boundary into the interior and this is done by allowing the points very near the 
boundary to average over pixels lying slightly outside. Two other considerations
enter. On the sphere the only efficient way to compute convolutions is by using
harmonic transforms. Therefore a small number of kernels is used whose radii
vary according to a geometric progression. Also the kernels profiles are chosen
to be annular with a finite width and solid for the very smallest in order to
avoid errors associated with Gibbs oscillations and inaccuracies of the spherical
harmonic transform on the scale of the pixelization. 
[See Fig.~\ref{LaplaceSolver:Fig}]

\subsection{Improvement of the solution near the boundary}
\label{ImpBound:Sect}

Starting with a map filled in using the Laplace solver, we improve the 
solution near the boundary of the masked region where the kernel profile 
becomes increasingly relevant. 
We define a series of strips $D_1\supset \ldots \supset D_n$ consisting
of pixels a distance $w_1>\ldots >w_n$ within the boundary, respectively,
and the noise is boosted by a factor of $\beta _1>\ldots >\beta _n=1.$ 
The condition number of the resulting linear equation 
depends on both the width of the strip and the noise boost $\beta .$
The value of the largest eigenvalue is very nearly one. This corresponds to
small-scale structure annihilated by $K$ so that $(I-K)$ acts very nearly
as the identity operator on small scales. Large scale configurations, on the other
hand, lead to small eigenvalues. 
The regularization of the smallest eigenvalues (so that they do not lie
too close to zero) can be interpreted as leakage from the boundary. For very 
large-scale structure, $K$ acts almost as the identity, because the integral over the 
the kernel is unity. However near the boundary, 
leakage takes place. On a long narrow strip the eigenfunction has approximately a 
sine function transverse profile, and the lowest eigenvalue (or equivalently the inverse
of the condition number) is of order $(d/w)^2$ where $d$ is the effective width 
of the kernel, which is fattened as $\beta $ is increased. Note that the effective
width may be much smaller than the actual support of the kernel because of oscillations of the
kernel as seen above.

A key input to choose the parameters of our method is
approximating the condition number
of the linear system resulting for a given power spectrum
and instrument noise pair as a function of strip width
and noise boost factor. We used a Lanczos method to
determine the condition number. The basic approach is
to generate a Krylov space in much the same way as
for the conjugate gradient method choosing a basis such that matrix 
$(p_i,Ap_j)_A$ is triadiagonal. Here the directions 
$p_i$ are A-conjugate. The range of the eigenvalues
of this subspace are supposed to be representative 
of the range for the full matrix $A.$
After a reasonable number of iterations, the maximum and 
minimum eigenvalues of the matrix, calculated using a
standard $QR$ algorithm, converge. 
(See \cite{GolubVanLoan} for a nice overview of Lanczos methods.)
A knowledge of the 
condition number is invaluable both for predicting
the rate of convergence and formulating a suitable stopping 
rule. 

We chose our sequence so that $\beta $ descends to one 
following more or less 
a geometric progression and $w$ is adjusted to obtain a condition number
in the neighborhood of $10^2.$ In this way 20 iterations per strip
of the conjugate gradient method were found to suffice. 
Our values of $(\beta ,w)$ where $(2000,100),$ $(400,35),$ $(40,15),$ $(1,7).$
Fig.~\ref{FillInHealFig} shows the final result starting on a strip of width 100 pixels.

The above method alone with the parameters chosen would work for holes and 
masks of width no larger than $1.5^\circ .$ But realistic Galactic masks are typically
much larger than this, up to $40^\circ $ wide or more. One option would be to 
start the sequence with additional strips with even larger noise boost factors and 
widths. We instead chose to capitalize on the fact that in the far interior the 
maximum likelihood solution very nearly solves the Laplace equation. 

\begin{figure}
\begin{center}
\includegraphics[width=4.9cm]{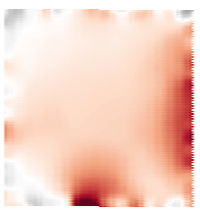}
\includegraphics[width=5cm]{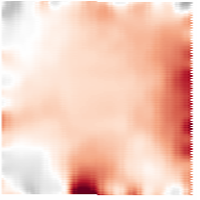}
\includegraphics[width=4.8cm]{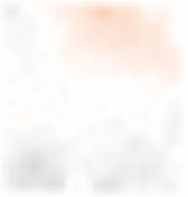}\\
\includegraphics[width=10cm]{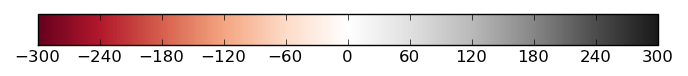}
\end{center}
\caption{
{\bf Validation of approximate solution.}
From left to right we show the approximate solution calculated according to the procedure described in Sects. 
\ref{DefineMask:Sect}-\ref{ImpBound:Sect},
the exact solution calculated using a direct method (i.e., Gaussian elimination) to solve the linear system
using the exact kernel on the entire domain, and the difference map. The size of the domain is $100^2$
pixels in the $n_{side}=2048$ Healpix pixelization.
}
\label{Validation:Fig}
\end{figure}

\subsection{Validation of above procedure}

We validated the procedure described in the previoius subsections by comparing to solution
obtained using a direct method (i.e., Gaussian elimination)
for solving the linear system 
$(({\bf C}^{-1})_{\bf xx}){\bf x}=(({\bf C}^{-1})_{\bf xy}){\bf y}$
where ${\bf x}$ denotes the vector of masked pixels. 
We used a square mask of 100 pixels on a side centered on the equator.
The direct method, whose computational complexity scales as ${n_{masked}}^3,$
is still feasible on a domain of this size and has the advantage that it can deal with the large condition 
number of the matrix $({\bf C}^{-1})_{\bf xx}$ without difficulty. 
In Fig.~\ref{Validation:Fig}, we compare the maximum likelihood solution using both methods. 
The agreement is quite good but not perfect. We have not fully explored the optimization of the 
parameters governing the improvement along the boundary of the masked region. 
As the strips become wider (entailing a larger condition number and hence requiring more iterations),
the approximate solution becomes increasingly accurate.
A study of the accuracy required for probes of primordial
non-Gaussianity as well as a comparison with competing methods of filling in, which unlike the present
one are non-Gaussian, will be addressed in a forthcoming publication [\cite{VanTent}].

\section{Concluding remarks}

We have demonstrated a workable filling in procedure for masked regions of arbitrary width
on the pixelized sphere, which we have validated using a Healpix pixelization of $n_{side}=2048.$ There
is no reason why the methods developed here should not extend to even finer 
pixelizations because the dimensionless ratios defining the numerical
difficulty of the problem do not involve the overall size of the sphere.
Moreover for the Laplace solver presented here the number of iterations needed
scales as a power of the logarithm of the ratio of two angular scales, allowing wide
masks to be treated with only a very modest increase in the number of iterations 
as will be necessary for very aggressive Galactic cuts. 

As explained in Section~\ref{Efficient:Sect}, the hardest part of realizing constrained 
Gaussian realizations is calculating the maximum likelihood for the missing pixels 
because the random component can be computed using the method introduced by Hoffman
once one has a working and efficient procedure for calculating the maximum likelihood
configuration. 

One of the disappointing conclusions of this study is that the 
preconditioned conjugate gradient method that we successfully implemented
on a comparable flat domain with periodic boundary conditions
did not work when carried over to the sphere.
The difference between the two cases is difficult to understand and plausibly
attributable to the difficulties of implementing a spherical harmonic transform
and its inverse as discussed in Section \ref{PixSphere:Sect}.

One interesting feature of the methods introduced in this paper
are that they are kernel based and thus to some extent independent
of the pixelization used. The basic operation is convolving maps 
with kernels, so even though the maps used are pixelized there
is no implicit finite difference scheme involved. The only requirement
is that the pixelization be fine enough so that the convolutions
calculated are sufficiently accurate. It is interesting that
the idealization of white instrument noise allows us to
analyze what is going on independent of any pixelization.

A code to carry out the procedures as described in the text is available from the authors 
upon request. 

\section{Discussion}

{\bf Acknowledgements:} We would like to thank 
Jean-Francois Cardoso, Joanna Dunkley, and 
Bartjan van Tent for useful discussions and 
Jacques Delabrouille and Guillaume Castex for providing
Galactic and point source masks for testing our code. 
The authors acknowledge the use of the Healpix package \cite{gorski} and
the healpy python interface (http://code.google.com/p/healpy/)
written by Cyrille Rosset for many of the computations carried out in this work.
.
TL acknowledge the use of Flipper code for flat sky maps analysis.

\end{document}